\documentclass[prb,twocolumn,showpacs,preprintnumbers,amsmath,amssymb]{revtex4}% Physical Review B

\usepackage{graphicx}% Include figure files
\graphicspath{{fig/}}
\usepackage{dcolumn}% Align table columns on decimal point
\usepackage{bm}% bold math
\usepackage{amsmath,amssymb}
\usepackage[T1]{fontenc}
\usepackage{textcomp}
\usepackage{color}
\begin{document}

\title{Spin Currents and Magnon Dynamics in Insulating Magnets
}

\author{Kouki Nakata,$^{1}$ Pascal Simon,$^2$ and Daniel Loss$^{1}$}

\affiliation{$^1$Department of Physics, University of Basel,   Klingelbergstrasse 82, CH-4056 Basel, Switzerland   \\
$^2$Laboratoire de Physique des Solides, CNRS UMR-8502, Universit$\acute{e} $ Paris Sud, 91405 Orsay Cedex, France  
}

\date{\today}

\begin{abstract}
% In contrast to the usual regular research paper, I tried to explain and specify the position of our work from the viewpoint of fundamental physics for the nonspecialist readers (e.g., elementary particle physics, strongly correlated spin physics, et al.); they willl judge whether they read by abstract and I believe this abstract motivates them. I guess the length of this abstract is OK for review articles; at least I love this and actually, this is the original my motivation to study spin transport.
Nambu-Goldstone theorem provides gapless modes to both relativistic and nonrelativistic systems.  The Nambu-Goldstone bosons in insulating magnets are called  magnons or spin-waves and play a key role in magnetization transport. We review here our past works on magnetization transport in insulating magnets and also add new insights, with a particular focus on magnon transport.
We  summarize in detail the magnon counterparts of electron transport, such as the Wiedemann-Franz law, the Onsager reciprocal relation between the Seebeck and Peltier coefficients, the  Hall effects, the superconducting state, the Josephson effects, and the persistent quantized current in a ring to list a few. 
%%%%%%%%%%%%%%%%%%%%%%%%%%%
%\ps{the following paragraph is completely obscure to me}. Kouki; simplified and revised.
Focusing on the electromagnetism of moving magnons, i.e., magnetic dipoles, we theoretically propose a way to directly measure magnon currents.
%Bosonic nature enables magnons to condensate and to form a macroscopic coherent state.
%Using an off-diagonal long-range order, we classify magnon states and reveal the resulting transport properties in details, with providing the corresponding spin-wave pictures.
%Taking into account the Bloch theorem, we theoretically exhibit the existence of persistent quantized magnon currents and propose a way to directly measure them based on the electromagnetism of magnetic dipoles.
%%%%%%%%%%%%%%%%%%%%%%%%%%%
As a consequence of  the Mermin-Wagner-Hohenberg theorem,  spin transport is drastically altered in one-dimensional antiferromagnetic (AF) spin-$1/2$ chains; where the N\'eel order is destroyed by quantum fluctuations and a quasiparticle magnon-like picture breaks down. Instead, the low-energy collective excitations of the AF spin chain are described  by a Tomonaga-Luttinger liquid (TLL) which provides the spin transport properties in such antiferromagnets some universal features at low enough temperature.
%%%%%%%%%%%%%%%%%%%%%%%%%%%
Finally, we enumerate open issues and provide a platform to discuss the future directions of magnonics. 
\end{abstract}
\pacs{Review article for special issue on magnonics from J. Phys. D.}
\maketitle

%%%%%%%%%%%%%%%%%
\section{Introduction}
\label{sec:Intro}
\begin{figure}[t]
\begin{center}
\includegraphics[width=8cm,clip]{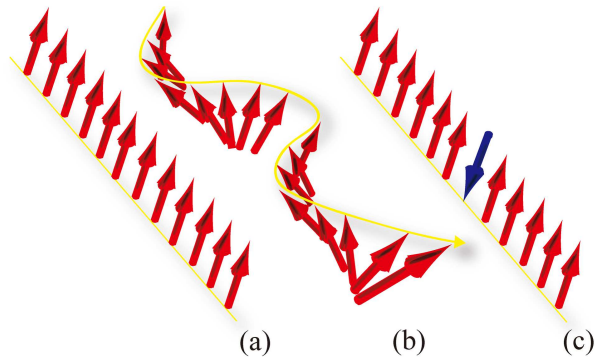}
\caption{(Color online)
Schematic representation of a spinwave, which characterizes a low-energy excitation in magnets. Its quantum counter part corresponds to a magnon excitation.
Such a Nambu-Goldstone boson appears by the spontaneous symmetry breaking [{\it{e.g.}}, $SO(3) \mapsto  SO(2)$ in (a)] and its dispersion vanishes in the long wavelength limit $k \rightarrow 0$ (i.e., a gapless collective mode).
Assuming the exchange interaction $J >0$ between the nearest neighboring spins of a ferromagnet with lattice constant $a$, 
we denote by $E_{\rm{a,b,c)}}$, the
energy of (a) the spin polarized state, (b) the spin-wave state, and (c) the spin-flip state, respectively. These energies can be estimated as follows;
$ \delta  E_{\rm{ba}} \equiv   E_{\rm{b}} - E_{\rm{a}} = {\cal{O}} (Ja^2k^2) \rightarrow  0 $ when $k \rightarrow  0$,
$ \delta  E_{\rm{ca}} \equiv   E_{\rm{c}} - E_{\rm{a}} = {\cal{O}} (J)$,
and 
$\delta  E_{\rm{ba}}/\delta  E_{\rm{ca}}= {\cal{O}} (a^2k^2) \ll  1 $ in the long wavelength limit.
\label{fig:spinwave} }
\end{center}
\end{figure}

%%%%%%%%%%%%%%%%%
% [Dear Daniel] Actually, this is the original my motivation to study spin transport.
Symmetry and its spontaneous breaking are fundamental concepts to understand and classify the low-energy physics of many systems.
%and unifying the underlying mechanism.
The Nambu-Goldstone theorem \cite{NG,NG2,NG3} ensures that when a continuous symmetry is spontaneously broken, gapless modes appear.
The massless \cite{NGboson,peskin,coleman,altland,Brauner} modes are called Nambu-Goldstone bosons and  their energy vanish in the long wavelength limit $k \rightarrow 0$.
The dispersion becomes linear in Lorentz-invariant (i.e., relativistic) systems, while it may be nonlinear in nonrelativistic systems and varies from one system to another (see Refs. [\onlinecite{HWng,HidakaNG}] ).
Well-known examples in solids are the so-called magnons \cite{BlochMagnon,HP} or spin-waves (Fig. \ref{fig:spinwave}).
On cubic lattices, the low-energy physics is characterized by nonrelativistic-like magnons with a quadratic dispersion in ferromagnets (FMs), while relativistic-like magnons emerge in antiferromagnets (AFs) where the N\'eel order is developed. 
%%%%%%%%%%%%%%%%%%%%%%%%%%%%%%%%%%%%%%%%%%%%%%%%%

The situation changes in lower dimensions ${\cal{D}} \leq 2$.
The Mermin-Wagner-Hohenberg theorem \cite{MW,Hohenberg} prohibits the spontaneous breaking of continuous symmetries at finite temperature in the thermodynamics limit (see Ref. [\onlinecite{DLFabioLeggett}] for an extension). Therefore,
spontaneous magnetic orders in ${\cal{D}} \leq 2$ spin systems are generally not possible at any finite temperature.
Indeed, the N\'eel order is destroyed by quantum fluctuations (i.e. strong quantum effects) in one-dimensional antiferromagnetic spin-$1/2$ chains
and the quasiparticle picture ({\it{e.g.}}, magnon) breaks down; instead the low-energy physics is described by the Tomonaga-Luttinger liquid \cite{TLL,TLL2} (TLL).
%%%%%%%%%%%%%%%%%%%%%%%%%%%%%%%%%%%%%%%%%%%%%%%%%

The dimensionality thus plays a crucial role on magnetism.
We summarize here spin transport properties in $\cal{D}$  dimensions, with a particular focus on magnons transport in insulating bulk magnets.
Quantum-statistical mechanics provides two kinds of particles; bosons and fermions. 
Electrons are fermions bounded by the Pauli exclusion principle, while magnons are bosons free from it. One may think at first sight
that this drastic difference in statical properties of these excitations will lead to strong differences between magnon and electron transport. In this paper,  we instead  describe many magnon counterparts of electron transport (see Table \ref{tab:analogue} for a comparison).
%Still, can magnon transport be similar to electron transport ?
%The answer is `YES'.
We review our past works on spin transport in insulating magnets and also add new insights to it \cite{LossPersistent,LossPersistent2,dipole,magnon2,Trauzettel,KevinPRL,KevinHighliht,KevinHallEffect,Kevin2,Kevin3,KKPD,KPD,magnonWF,QHEmagnon}.

The bosonic nature of magnons enables, however, excited magnons to dynamically condensate \cite{demokritov}.
We then reveal the resulting transport properties \cite{KKPD,KPD} analogous to supercurrents of  superconductors and summarize our recent results on universal thermomagnetic relations \cite{magnonWF} in insulating magnets.
We also provide a platform to discuss the next direction of magnon spintronics, dubbed magnonics \cite{MagnonSpintronics,ReviewPRapply,magnonics,spincal,spincalreview,caloritronics}, based on fundamental physics
(see Refs. [\onlinecite{MagnonSpintronics,ReviewPRapply}] for some applications).

%%%%%%%%%%%%%%%%%%%%%%%%%%%%%%%
\begin{table*}
\caption{
\label{tab:analogue}
Magnon analogues of electron transport.
Spin Seebeck \cite{uchidametal,uchidainsulator,phonon} and Peltier \cite{Peltier} effects in a metal/FI junction were observed by the inverse spin Hall effect \cite{ISHE1},
and the Onsager relation between them and the deviation were observed in Ref. [\onlinecite{OnsagerExperiment}].
Ref. [\onlinecite{onose}] reported the observation of magnonic thermal Hall effect \cite{katsura}
and magnonic Snell's law has been experimentally established in Ref. [\onlinecite{Snell_Exp}]. 
See Ref. [\onlinecite{WFchain}] for a WF law of spin transport in one-dimensional integrable spin-$1/2$ $XXZ$ chain \cite{SF}
and Refs. [\onlinecite{WFnFL,WFnFL2,WFnFL3,WFnFL4,WFnFL5}] for the progress\cite{FQHEnote} on the WF laws of non-Fermi liquids.
A spinon spin current in one-dimensional spin-$1/2$ chain has been observed in Ref. [\onlinecite{SpinonCurrent}].
Refs. [\onlinecite{NernstAF,NernstAF2}]  have theoretically proposed magnon spin Nernst effects in antiferromagnets\cite{AFspintronics,SekiAF,DL_rev_QuantSpinDeco}.
}
%%%%%%%%%%%%
\begin{ruledtabular}
\begin{tabular}{cccccc}
%%%%%%%%%%%%
    Electron transport \cite{kittel} $e$ &  Magnon transport  $\mu_{\rm{m}}$ \\ \hline
 Electric current  &     Spin-wave spin current \cite{spinwave}    \\
  Seebeck effect   &     Spin Seebeck effect \cite{uchidametal,uchidainsulator,phonon,UchidaReview2016,adachi,adachiphonon,adachireport,xiao,silas} \\
    Peltier effect   &     Spin Peltier effect \cite{Peltier,YTpeltier}  \\
Wiedemann-Franz law \cite{WFgermany}     & Magnonic Wiedemann-Franz law  \cite{magnonWF}  \\
Superconductive state    & Quasiequilibrium magnon condensate \cite{demokritov,BECcurrentNOTE}  \\
 Josephson effect \cite{Josephson}  &  Magnon Josephson effect \cite{KKPD} \\
 Persistent current:   & `Persistent' current \cite{KKPD} of quasiequilibrium condensates:  \\
   Quantization  in superconducting ring (i.e., fluxoid)  &   Quantization  in  magnon condensate ring \cite{KKPD}  \\
  Electronic  quantum RC circuit \cite{RC,RC2,RC3,RC4,RC5}  &   Magnetic   quantum RC circuit \cite{KevinPRL}\\
    Electronic  transistor  &  Magnon transistor \cite{KevinHighliht,TserkovnyakNatNano} \\
\end{tabular}
\end{ruledtabular}
\end{table*}
%%%%%%%%%%%%%%%%%%%%%%%%%%%%%%%%%%$^{3}\rm{He}$

The paper is organized as follows.
In Sec. \ref{sec:1dim}, considering quasi-one-dimensional spin chains of  finite size, we describe the spin transport carried by magnons in FMs and the one by TLL in AFs, and reveal the intrinsic properties to one-dimensional AFs.
In Sec. \ref{sec:Hall}, introducing a Berry phase particular to magnetic dipoles, we demonstrate that it generates phenomena analogous to Hall effects in two-dimensional magnets [{\it{e.g.}}, ferromagnetic insulators (FIs)].
In Sec. \ref{sec:3dim}, focusing on a three-dimensional ferromagnetic insulating junction, we discuss the transport properties of both non-condensed and condensed magnons, and clarify the differences;
after reviewing fundamentals of magnons in Sec. \ref{subsec:3dim}, we consider the magnon and heat transport, and find the universal thermomagnetic relation in Sec. \ref{subsec:WF}. The magnon states are classified in detail and the resulting transport properties, represented by Josephson effects {\it{etc}}., are summarized in Sec. \ref{subsec:Josephson}.
The persistent current and the quantization are discussed in Sec. \ref{subsec:persistent}, and we theoretically propose how to directly measure it in Sec. \ref{subsec:Measurement} based on the electromagnetism of magnetic dipoles.
Finally, we summarize in Sec. \ref{sec:remark} and enumerate open issues  in Sec. \ref{sec:outlook}.

%%%%%%%%%%%%%%%%%%%%%%%%%%%
\section{Spin transport in  spin chains}
\label{sec:1dim}
%%%%%%%%%%%%%%%%%%%%%%%%%%%

The Mermin-Wagner-Hohenberg theorem \cite{MW,Hohenberg,DLFabioLeggett} prohibits the spontaneous magnetic order in ${\cal{D}} \leq 2$ isotropic spin systems at finite temperature (with sufficiently short-ranged exchange coupling), but only in the thermodynamic limit; effective ordering in nanostructures of a finite size at sufficiently low temperatures is possible.
We then consider a quasi-one-dimensional spin chain of a finite length and investigate the magnetization transport both in FMs and AFs.

%%%%%%%%%%%%%%%%%%%%%%%%%%%
\subsection{Spin Conductances}
\label{subsec:Gs}
%%%%%%%%%%%%%%%%%%%%%%%%%%%

\begin{figure}[h]
\begin{center}
\includegraphics[width=8cm,clip]{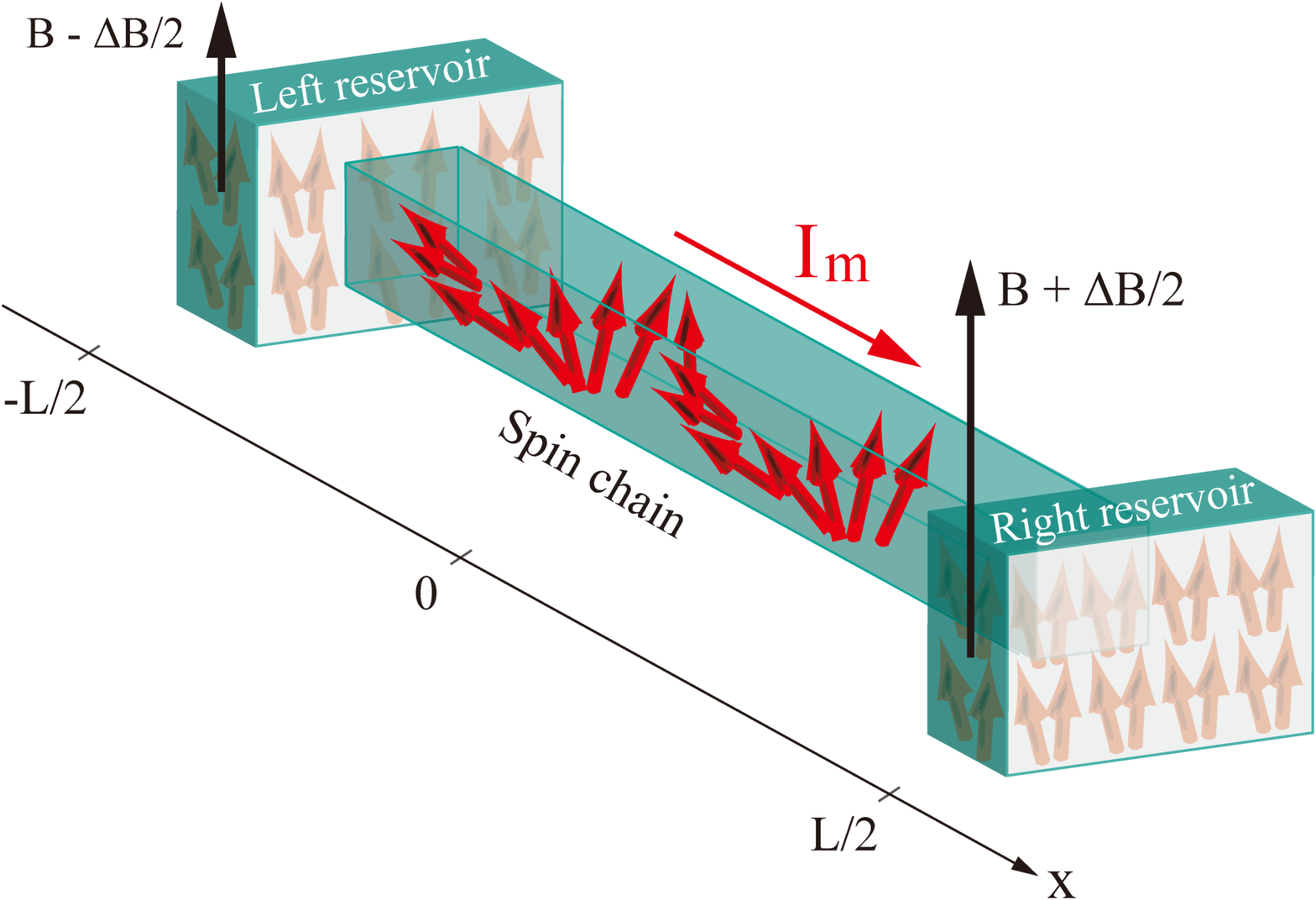}
\caption{(Color online)
Schematic representation of a spin chain with a finite length $L$ sandwiched between two large three-dimensional magnetic reservoirs. 
A small spatially varying magnetic field [Eq. (\ref{eqn:deltaB})]  is superimposed on the offset field $ B {\mathbf{e}}_z$, which interpolates smoothly between the values $\pm  \Delta B/2$ in the reservoirs.
The field gradient produces a magnetization current $I_{\rm{m}}=  G_S \Delta B$.
The setup was proposed in Ref. [\onlinecite{magnon2}]. 
%{\color{red} The figure here is different, so we do not need permission} and reprinted with permission.
\label{fig:chain} }
\end{center}
\end{figure}

We consider \cite{magnon2} a quasi-one-dimensional spin system illustrated in Fig. \ref{fig:chain}.
The spin chain of finite length $L$ is sandwiched between two bulk (i.e., large three-dimensional) magnets which work as reservoirs for magnetization.
The reservoirs narrow adiabatically around the transition regions $ \mid  x  \mid  \approx  L/2  $ and are connected to the chain.
A small spatially varying magnetic field $ \delta  B (x) {\mathbf{e}}_z$  is superimposed on the offset field $ B {\mathbf{e}}_z$ with (Fig. \ref{fig:chain})
\begin{eqnarray}
  \delta  B (x) = 
\begin{cases}
- \Delta B/2
&   {\textrm{ for }}     \     x < -L/2,
  \\ 
%%%%%%%%%%%%%%%%%%%
 \Delta B/2
&     {\textrm{ for }}    \  x > L/2,
\end{cases}
\label{eqn:deltaB} 
\end{eqnarray}
which interpolates smoothly through the region $ \mid x \mid  <L/2 $ between the values $\pm  \Delta B/2$ in the reservoirs (see Ref. [\onlinecite{magnon2}] for details). These extra-fields $\pm  \Delta B/2$ act like chemical potentials and need to be kept constant
by a spin battery and/or by spin-lattice relaxation processes which keep the magnon occupation number at a constant value corresponding to the lattice temperature and Zeeman energy in each reservoir. Thus, a chemical potential difference acts then like a force on the magnons and drives a current of magnons. The flow of magnons from one reservoir to the other should be slow compared to the time it takes to refill the leaving/arriving magnons  in each reservoir.
The field gradient (acting like a chemical potential gradient) produces a magnetization current $I_{\rm{m}}$ from the left to the right reservoir
\begin{eqnarray}
  I_{\rm{m}} =  G_S \Delta B,
\label{eqn:Imdef} 
\end{eqnarray}
where  $G_S$ is the spin conductance; like for electrons \cite{Datta}, it remains finite  in the ballistic limit due to the contact resistance between the reservoirs and the chain \cite{magnon2}.

The spin chain extends from $x = -L/2$ to $L/2$ and it is described by
\begin{eqnarray}
H_{\rm{chain}}  = J \sum_{\langle i j \rangle} {\mathbf{S}}_{i} \cdot  {\mathbf{S}}_{j} + g \mu _{\rm{B}}\sum_{i}B_i S_i ^{z},
\label{eqn:H} 
\end{eqnarray}
where $ J < 0 $ is the exchange interaction between the nearest neighbor spins ${\mathbf{S}}_{i}$ and ${\mathbf{S}}_{j}$ on sites $i, j$
for FMs and $J > 0$ for AFs.
The magnetization is characterized by \cite{HP,Mattis} magnons in a FM, while for spin-$1/2$ AF chain it can be mapped to spinless fermions via Jordan-Wigner transformation which can then be bosonized to give rise Tomonaga Luttinger Liquids (TLLs) \cite{giamarchi,HaldaneTLL,AffleckTLL,Fradkin}.

%%%%%%%%%%%%%%%%%%\cite{TLL,TLL2}
We arrange the spin chains in parallel.
Assuming a weak intermediate exchange interaction $J_{\rm{inter}}$  between the spin chains $ |J_{\rm{inter}}  |  \ll  | J  | $,
they can be regarded as being uncoupled with each other at temperatures $|  J_{\rm{inter}} |   \ll   k_{\rm{B}} T   \ll  |  J  | $.
Linear response theory provides the spin conductances (see Ref. [{\onlinecite{magnon2}}] for details)
\begin{eqnarray}
    G_S =  N_{\rm{chain}} \frac{(g \mu _{\rm{B}})^2}{h} 
\begin{cases}
n_{\rm{B}}(g \mu _{\rm{B}}B)
&     {\textrm{ for }}    \    {\textrm{FM}}, 
  \\ 
 1/K_b   
&   {\textrm{ for }}     \      {\textrm{AF}},
\end{cases}
\end{eqnarray}
where $N_{\rm{chain}}$ is the number of spin chains, the Bose distribution function $n_{\rm{B}}(\epsilon) = 1/[{\rm{exp}}(\epsilon/k_{\rm{B}}T)-1] $, and the interaction parameter $K_b$ in the reservoirs; it becomes $K_b  \simeq  4\sqrt{3}/\pi $ for isotropic antiferromagnetic bulk reservoirs (and $K_b = 1$ for an XY AF).

%%%%%%%%%%%%%%%%%%%%%%%%%%%%%%%%%%%%%%%%%%%%%%%
\begin{table*}
\caption{
\label{tab:BerryPhase}
Geometric phases for electrons of charge $e$ and magnons of magnetic moment $\mu_{\rm{m}}$ moving along a path $\gamma $ for the geometry under consideration.
Both are special cases of the Berry phase \cite{Mignani,Hea}.
Moving magnons in electric fields ${\mathbf{E}} $ acquire the A-C phase $\theta _{\rm{A{\mathchar`-}C}} $, 
which is analogous to the A-B phase \cite{DL_rev_QuantSpinDeco,DL_QuantCoherence} $\theta _{\rm{A{\mathchar`-}B}} $ for electrons.
The Dzyaloshinskii-Moriya (DM) interaction \cite{DM,DM2,DM3,onose,ACspinwave} acts as an electric vector potential and can be identified \cite{katsura2,Mook2,Lifa} with the A-C effect. 
}
%%%%%%%%%%%%
\begin{ruledtabular}
\begin{tabular}{cccccc}
%%%%%%%%%%%%%%%%%%%%%%%%%%%%%%%%%%%%%%%%%%%%
    Aharonov-Bohm (A-B) phase \cite{bohm} &  Aharonov-Casher (A-C) phase \cite{casher}   \\ \hline
%%%%%%%%%%%%%%%%%%%%%%%%%%%%%%%%%%%%%%%%%%%%
 Particle with electric charge  $e $     &    Magnon with magnetic dipole moment $ {\mathbf{\mu}}_{\rm{m}}= - g\mu_{\rm{B}}{\mathbf{e}}_z   $ \\
  Magnetic vector potential $ {\mathbf{A}}  $  &  Electric `vector potential' $ {\mathbf{A}}_{\rm{m}} \equiv  {\mathbf{E}} \times  {\mathbf{\mu}}_{\rm{m}}  $\\
 $ \theta _{\rm{A{\mathchar`-}B}} = [e/{(\hbar c)}] \int_{\gamma} \textrm{d} {\bf l} \cdot  {\bf A}$   &  
$ \theta _{\rm{A{\mathchar`-}C}} = [{g \mu_{\rm{B}}}/{(\hbar c^2)}] \int_{\gamma} \textrm{d} {\bf l} \cdot   ({\bf E} \times  {\bf e}_z) $ \\
\end{tabular}
\end{ruledtabular}
\end{table*}
%%%%%%%%%%%%%%%%%%%%%%%%%%%%%%%%%%

%%%%%%%%%%%%%%%%%%%%%%%%%%%
\subsection{Power Dissipation}
%%%%%%%%%%%%%%%%%%%%%%%%%%%

Assuming that the spin chain $H_{\rm{chain}}$ [Eq. (\ref{eqn:H})] and the magnetic reservoirs in Fig. \ref{fig:chain} consist of AFs ($J >0$), 
we \cite{Trauzettel} generalized the above theory to the response to an ac magnetization source where the magnetic field bias $\Delta  B$ (Fig. \ref{fig:chain}) changes periodically in time at a given driving frequency $\omega $: Eq. (\ref{eqn:deltaB}) is replaced by 
\begin{eqnarray}
  \delta  B (x) = 
\begin{cases}
- (\Delta B/2) {\rm{cos}}(\omega t)
&   {\textrm{ for }}     \     x < -L/2,
  \\ 
 (\Delta B/2)  {\rm{cos}}(\omega t)
&     {\textrm{ for }}    \  x > L/2,
\end{cases}
\label{eqn:deltaBac} 
\end{eqnarray}
and the oscillating magnetic field produces the magnetization current $I_{\rm{m}}$ in a $XXZ$ spin chain described by
\begin{eqnarray}
H_{XXZ}  =  J \sum_{\langle i, j \rangle} (S_{i}^{x} S_j^{x} +  S_{i}^{y} S_j^{y} + \Delta  S_{i}^{z} S_j^{z}),
\end{eqnarray}
where we assume \cite{giamarchi}
%$J,\Delta  >0$ 
$ 0 < J $ and $0 < \Delta < 1$
namely antiferromagnetic interactions.
%%%%%%%%%%%%%%%%%
Again, 
%After a Jordan-Wigner transformation and using standard bosonization techniques,\cite{giamarchi,Fradkin,HaldaneTLL,AffleckTLL}  
the low-energy effective theory of the spin Hamiltonian  $H_{XXZ} $ is given by a TLL,
\begin{eqnarray}
H_{\rm{TLL}}  =  \frac{\hbar  v }{2} \int d x  \Big[ g_{\rm{TLL}} [\Pi (x)]^2 +  [\partial _x \varphi  (x)]^2/ g_{\rm{TLL}}    \Big],
\label{eqn:htll}
\end{eqnarray}
where $v$ is the velocity of the  spinon excitation and $g_{\rm{TLL}}$ denotes the Luttinger interaction parameter given by \cite{giamarchi} $g_{\rm{TLL}}^{-1}= ({2}/{\pi})\arccos(-\Delta)$; it becomes $g_{\rm{TLL}} \approx  (1 + 4 \Delta /\pi)^{-1/2}$ at $0 < \Delta\ll 1$.
We also introduced the Bose field operator  $\varphi  (x)$ in the bosonization language associated with spinon excitations and its conjugate momentum density  $\Pi (x)$. Note that we have ignored Umklapp scattering in Eq. (\ref{eqn:htll}). A non-interacting system corresponds to 
$g_{\rm{TLL}}=1$ (i.e. $\Delta=0$).
%%%%%%%%%
The reservoirs can also be described within this formalism by introducing an inhomogeneous TLL. This amounts to 
assign a spatial dependence
to $v$ and $g_{\rm{TLL}}$ such that $v(x)=v_r$ and $g_{\rm{TLL}}(x)=g_r$ in the reservoirs (for $|x|>L/2$) and 
$v(x)=v_c$, $g_{\rm{TLL}}(x)=g_c$  in the spin chain region (for $|x|<L/2$). We typically expect $g_r\approx 1$.
Within this formalism, 
the magnetization current $I_{\rm{m}}(\omega )$ and the spin conductance $G_S$ [Eq. (\ref{eqn:Imdef})] can be evaluated using linear response theory (see Refs. [\onlinecite{magnon2,Trauzettel}] for details).
%We find \cite{{Trauzettel} that  $I_{\rm{m}}(\omega )$ can be expressed as

%%%%%%%%%
Then the magnetic power $W_{\rm{m}}(\omega )$ is defined by $W_{\rm{m}} (\omega ) \equiv   I_{\rm{m}}^2(\omega )/(2G_S) $
in analogy to the electric power (i.e., Joule heating)  $W_{\rm{e}} \equiv   I_{\rm{e}} ^2/(2G_{\rm{e}})$, 
where $I_{\rm{e}} $ is an electric current, $G_{\rm{e}} $ the conductance , and $\Delta  V$ an external voltage difference.
The finite frequency absorption power can be expressed as \cite{Trauzettel}
\begin{eqnarray}
W_{\rm{m}}(\omega) &=& g_c \frac{(g \mu_B \Delta B)^2}{2 h} \left( \frac{\sin
({\tilde{\omega}}/2)}{{\tilde{\omega}}/2} \right)^2   \label{eq:W}  \\
&\times& \frac{1-\gamma^4 + 2 \gamma (1-\gamma^2)
\cos({\tilde{\omega}})}{1+\gamma^4-2\gamma^2
\cos(2{\tilde{\omega}})} , \nonumber 
\end{eqnarray}
where $\gamma = (g_r-g_c)/(g_r+g_c)$
is the reflection coefficient of spinon excitations at the  sharp boundary between the chain and its reservoirs and
$\tilde{\omega}=\omega L/v_c$ the ratio between the frequency and  level spacing in the finite size chain.
Note in passing that a measurement of the absorbed power due to ac
excitation of the quantum spin chain provides a  way to
measure interaction dependent coefficients such as $g_c$.
Another conclusion can be drawn from  Eq. (\ref{eq:W}): we notice that $W_{\rm{m}}(\omega)$ vanishes as $(\sin ({\tilde{\omega}}/2))^2$ close to $\omega\approx  2 \pi n L/ v_c$ with $n$ integer. One can show instead that the spin current resulting from a
 continuous wave radiation vanishes only as $\sin ({\tilde{\omega}}/2)$ close to that driving frequency. This implies that
the magnetic power absorption is more strongly suppressed than the magnetization current at frequencies close to $2 \pi n L/v_c$. This feature could be used to transfer a spin current (thus data) at special frequencies with low power dissipation.

In the limit $\omega  \rightarrow  0$, Eq. (\ref{eq:W}) reads
\begin{equation}
W_{\rm{m}} = g_r(g \mu _{\rm{B}} \Delta  B)^2/(2 h) ,
\end{equation}
to be compared with the typical electric power $W_{\rm{e}} = (e \Delta  V)^2/h $.
Let us now give estimate of these two powers consumptions.
For typical values $\Delta V = 1$mV and $\Delta  B =0.1$T, we find
\begin{eqnarray}
  W_{\rm{m}} =  {\cal{O}}(10^{-15}) {\rm{J}}/s  \approx   10^{-4}    W_{\rm{e}}, 
\end{eqnarray}
where $g_r=1$ was assumed.
We can thus conclude that substantial advantages with respect to power consumption can be found  using spin magnetization currents in insulators instead of 
electric currents.

\begin{figure}[h]
\begin{center}
\includegraphics[width=8.5cm,clip]{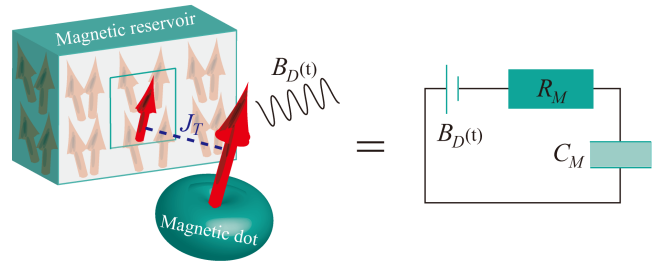}
\caption{(Color online)
Schematic representation of a quantum magnetic RC circuit proposed in Ref. [\onlinecite{KevinPRL}],
which consists of the magnetic resistance $R_M$, the capacitance $C_M$, and the applied magnetic field $B_D$.
The magnetic dot and the reservoir are weakly coupled by the exchange interaction $J_T $ and both of them are modeled as one-dimensional chains.
%Reprinted with permission from Ref. [\onlinecite{KevinPRL}].
\label{fig:QD} }
\end{center}
\end{figure}

%%%%%%%%%%%%%%%%%%%%%%%%%%%
\subsection{Quantum Magnetic RC Circuit}
\label{subsec:RC}
%%%%%%%%%%%%%%%%%%%%%%%%%%%
 By analogy with  quantum optics where the single-photon source is a major element to encode or manipulate a quantum state~\cite{Gisin2002}, or with quantum electronics where an on-demand electron source has been recently realized~\cite{Feve2007,Mahe2010,Dubois2013}, we have recently proposed how to realize some on-demand single magnon or spinon excitations using magnetic insulators.\cite{KevinPRL}

In analogy to the charged-based quantum RC circuit \cite{RC,RC2,RC3,RC4,RC5} (Table \ref{tab:analogue}), we proposed a quantum magnetic RC circuit as depicted in Fig. \ref{fig:QD} that  could potentially act as an 
on-demand coherent source of magnons or spinons.\cite{KevinPRL}
In Fig. \ref{fig:QD} the magnetic dot is weakly exchange coupled to a large magnetic reservoir and both of them are assumed to be nonitinerant magnets.  We describe these insulating magnets by a Heisenberg Hamiltonian for one-dimensional spin chains (see Ref. [\onlinecite{KevinPRL}] for details).
%%%%%%%%%%%%%%%%%%%%%%
A static-and time-dependent component of magnetic field $B_D(\omega )$ is applied to the dot
and the excess magnetization of the magnetic dot $M_D(\omega )$ is characterized by
\begin{eqnarray}
  \frac{M_D (\omega )}{B_D (\omega )} = C_M (1 + i \omega  C_M R_M),
\label{eqn:RC} 
\end{eqnarray}
where $R_M$  and $C_M$ are the magnetic resistance and the capacitance respectively of the equivalent RC circuit (see Fig.
\ref{fig:QD}). The excess magnetization is defined as $M_D(\omega) = g\mu_B N_D(\omega)$, where $N_D(\omega)$  is the Fourier transform of  the
time-dependent  excess number of magnetic quantum excitations (magnons or spinons) in the dot magnetic insulator.
We describe both the magnetic dot and the reservoir by spin chains. They can be both modeled by the spin anisotropic  Heisenberg Hamiltonian. The magnetic dot Heisenberg Hamiltonian reads
\begin{equation}
H_D = \sum_{\langle i j\rangle} {\bf S}_i \cdot {\bf J}_D \cdot {\bf S}_j +  g\mu_B \sum_i  {\bf B}_D(t) \cdot {\bf S}_i.
\label{eq:Ham}
\end{equation}
${\bf J}_D$ denotes a diagonal $3\times 3$-matrix with $\textrm{diag}({\bf J}_D) = J_D\{ 1, 1, \Delta_D\}$. $J_D$ is the magnitude of the exchange interaction and $\Delta_D$ the anisotropy in our model. $J_D \lessgtr 0$ corresponds respectively to the FM and the AF ground state. The field ${\bf B}_D(t) = B_D(t) {\bf e}_z$ denotes the time-dependent magnetic field applied to the dot. A similar Hamiltonian can be used to describe the reservoir (introducing different coupling constants) with only a constant magnetic field.

We couple these two systems, the  magnetic dot and the reservoir, via some
exchange interaction  of the form $ J_T {\mathbf{S}}_R \cdot  {\mathbf{S}}_D $,
where ${\mathbf{S}}_R$ is the last spin in the reservoir and  $ {\mathbf{S}}_D $ the first spin in the dot.

Using the linear response theory, we can express the change in magnetization $M_D(\omega)$ due to a small time-dependent change in $B_D(\omega)$ in terms of the  density retarded Green function, $M_D(\omega) = (g\mu_B)^2G_\textrm{ret}(\omega) B_D(\omega)$ where 
$G(\omega) = i \int_0^\infty d t e^{i\omega t} \langle \hat{N}_D(t)\hat{N}_D(0)\rangle_H$. Such correlation functions can be calculated order by order in perturbation theory in $J_T$ up to $O(J_T)^4$ (see [\onlinecite{KevinPRL}] for details), the results for $G(\omega)$ being identical for both ferromagnetic  and antiferromagnetic  systems.

We found that the spin resistance of AFs becomes {\it universal},
\begin{eqnarray}
  R_M=  \frac{h}{p (g \mu _{\rm{B}})^2 },
\label{eqn:RC2} 
\end{eqnarray}
with $p=2$ for a small magnetic dot (in the sense that the level spacing is always larger than the temperature) and $p=1$ for a large quantum dot.\cite{KevinPRL}  This implies that the resistance 
does not depend on material parameters of the magnetic dot. This result should be contrasted with the ferromagnetic case where the resistance is found to be generically non universal.\cite{KevinPRL}
These predictions can be tested either in cold atomic system where spin chains Hamiltonians have been realized or in chains of adatoms adsorbed on an insulating substrate. We refer the reader to
[\onlinecite{KevinPRL}] for a detailed discussion of possible experimental realizations.
%{\color{red}
%Discuss briefly:
%Ultrafast magnon-transistor at room temperature
%Kevin A. van Hoogdalem and Daniel Loss.
%Phys. Rev. B 88, 024420 (2013); arXiv:1209.5594; See News and Views, SPINTRONICS: An insulator-based transistor, by Yaroslav Tserkovnyak; Nature Nanotechnology 8, 706 (2013). 

%Frequency dependent transport through a spin chain
%Kevin A. van Hoogdalem and Daniel Loss.
%Phys. Rev. B 85, 054413 (2012); arXiv:1111.4803.

%Rectification of spin currents in spin chains
%Kevin A. van Hoogdalem and Daniel Loss.
%Phys. Rev. B 84, 024402 (2011) }
%%%%%%%%%%%%%%%%%%%%%%%%%%%%%%%%%%
\subsection{Rectification Effects and Magnon Transistor}
\label{subsec:KevinWork}
%%%%%%%%%%%%%%%%%%%%%%%%%%%%%%%%%%

Focusing again on both ferromagnetic and antiferromagnetic nonitinerant spin chains adiabatically connected to two spin reservoirs \cite{magnon2},
we \cite{Kevin2} have studied rectification effects.
Using spin-wave formalism (i.e., Holstein-Primakoff transformation\cite{BlochMagnon,HP}) and the Landauer-B$\rm{\ddot{u}}$ttiker approach \cite{kittel,Landauertext,Lesovik,Datta}, it has been found in Ref. [\onlinecite{Kevin2}] that a spin anisotropy combined with an offset magnetic field is the crucial ingredient to achieve a nonzero rectification effect of spin currents in FMs.
However, for AFs, a uniform anisotropy is not sufficient to achieve a sizable current rectification.
Using Jordan-Wigner transformation (with a bosonization procedure) and TLL formalism\cite{giamarchi,HaldaneTLL,AffleckTLL,Fradkin,TLL,TLL2},
we have studied the scaling behavior of the antiferrromagnetic spin chain; the renormalization-group analysis \cite{giamarchi} has shown that a spatially varying anisotropy, attained by a site impurity, instead realizes a sizable rectification effect.
%%%%%%%%%%%%%%%%%%%%%%%%%%
On top of this, using similar approach with the help of Schwinger-Keldysh formalism\cite{Schwinger,Schwinger2,Keldysh,rammer,haug,kamenev,tatara,kita,new},
we \cite{Kevin3} have studied frequency-dependent spin transport and proposed a system that behaves as a {\it capacitor} for the spin degree of freedom;
an anisotropy in the exchange interaction plays the key role in such a spin capacitor.
%%%%%%%%%%%%%%%%%%%%%%%%%%

%Ultrafast magnon transistor at room temperature
Finally, within a sequential tunneling approach for the describing the spin transport, 
we \cite{KevinHighliht} have proposed an ultrafast magnon transistor at room temperature and a way to combine three magnon transistors to form a  purely magnetic NAND gate\cite{NANDnote} in Ref. \onlinecite{KevinHighliht} (see also Ref. \onlinecite{TserkovnyakNatNano}).
Focusing on transport of magnons and spinons through a triangular molecular magnet\cite{MolecularMagnetText,MolecularMagnet,MolecularMagnet2,MolecularMagnet3,MolecularMagnet4},
it has been shown that electromagnetically changing the state of the molecular magnet, 
the magnitude of the spin current can be efficiently controlled.

%%%%%%%%%%%%%%%%%%%%%%%%%%%
\section{Magnon Hall Effects in Aharonov-Casher Phase}
\label{sec:Hall}
%%%%%%%%%%%%%%%%%%%%%%%%%%%

Since magnons are magnetic dipoles $ {\mathbf{\mu}}_{\rm{m}}= - g\mu_{\rm{B}}{\mathbf{e}}_z   $
%$ - g \mu _{\rm{B}}{\mathbf{e}}_z$, 
a moving magnon in an electric field ${\mathbf{E}}$ acquires an Aharonov-Casher (A-C) phase $\theta _{ij}$  (Table \ref{tab:BerryPhase}).
In a two-dimensional Heisenberg FM, this motion is described by a  spin (pointing along $z$) `hopping' from site ${{\mathbf{x}}_i}$ to a neighboring site 
${{\mathbf{x}}_j} $ and thereby picking up a
Peierls phase factor $e^{-i \theta _{ij}}$ with A-C phase 
$\theta _{ij} \equiv [g \mu _{\rm{B}}/(\hbar c^2)] \int_{{\mathbf{x}}_i}^{{\mathbf{x}}_j} d  {\mathbf{x}} \cdot  ({\mathbf{E}}\times {\mathbf{e}}_z) $, where the integration runs along the straight line connecting the two neighboring lattice sites ${{\mathbf{x}}_i}$ and ${{\mathbf{x}}_j}$.\cite{magnon2}
The corresponding Hamiltonian is given by \cite{magnon2} ($J < 0$)
\begin{eqnarray}
H_2  &=& \frac{J}{2} \sum_{\langle i, j \rangle} [S_i ^{+}S_j^{-}  {\rm{e}}^{-i \theta _{ij}}+S_i ^{-}S_j^{+}  {\rm{e}}^{i \theta _{ij}} 
+ 2S_i ^{z}S_j^{z}]  \nonumber \\
&+& g \mu _{\rm{B}}\sum_{i}B_i S_i ^{z},
\label{eqn:H_2} 
\end{eqnarray}
where the raising (lowering) operators $S_i ^{\pm }= S_i^x \pm i S_i^y $  effectively describe the hopping of the spin component $S_i ^{z}$.
%$\theta _{ij} \equiv [g \mu _{\rm{B}}/(\hbar c^2)] \int_{{\mathbf{x}}_i}^{{\mathbf{x}}_j} d  {\mathbf{x}} \cdot  ({\mathbf{E}}\times {\mathbf{e}}_z) $, where the integration is along a straight line connecting two nearest neighbor lattice sites.
The Holstein-Primakoff transformation \cite{HP} maps spins into the magnon degrees of freedom and  this Hamiltonian into
the single magnon Hamiltonian $ {h}_2$ in the presence of an effective 
electric `vector potential' $ {\mathbf{A}}_{\rm{m}} \equiv  {\mathbf{E}} \times  {\mathbf{\mu}}_{\rm{m}}  $,  \cite{magnon2}
\begin{eqnarray}
 {h}_2 = \frac{\mid J \mid S a^2 }{\hbar ^2 } ({\mathbf{p}} -  g \mu _{\rm{B}}{\mathbf{E}} \times  {\mathbf{e}}_z/c^2)^2 + g \mu _{\rm{B}}B,
 \label{eqn:h_2} 
\end{eqnarray}
where $a$ is the lattice constant. Then the mass of a magnon $m$ is defined by $  1/m \equiv  2 \mid J \mid S a^2/\hbar ^2   $.
The Heisenberg equation of motion,
$ m \ddot{\mathbf{x}} =  i ^2 [{h}_2, [{h}_2, {\mathbf{x}}]] \equiv   {\mathbf{F}}$, provides the force acting on a magnon, 
%\cite{magnon2} 
\begin{eqnarray}
 {\mathbf{F}} = - g \mu _{\rm{B}} {\mathbf{\nabla }}[B-({\mathbf{v}} \times {\mathbf{E}})\cdot {\mathbf{e}}_z/c^2].
 \label{eqn:F_2} 
\end{eqnarray}
This means that driven by a magnetic field gradient ${\mathbf{\nabla }} B$, magnons in an inhomogeneous electric field ${\mathbf{E}}({\mathbf{x}})$
experience a force ${\mathbf{F}}$ analogous to the Lorentz force, which leads to phenomena analogous to classical Hall effects \cite{magnon2}.
Thus the A-C effect gives a handle \cite{KKPD} to electromagnetically control magnon transport (see also Sec. \ref{subsec:Josephson}).
Recently, such an effect on magnons has been experimentally observed in Ref. [\onlinecite{ACspinwave}]
and magnonic Snell's law at interfaces, implying specular (elastic) reflection at the boundary to vacuum, has been experimentally established in Ref. [\onlinecite{Snell_Exp}]. 
%%%%%%%%%%%%%%%%%%%%%%%%%%%%%%%%%
Therefore, this analogy leads to the whole phenomenology of Hall effects such as the magnonic `quantum' Hall effect\cite{MagnonicQHEnote,QHEmagnon,Haldane2},  
the thermal Hall effect in skyrmion lattices\cite{KevinHallEffect}, 
the Hall effect in frustrated magnets\cite{Fujimoto}, and the magnonic topological insulators\cite{Matsumoto,Matsumoto2,RShindou,RShindou2,RShindou3,Mook,Mook2,Mook3,Lifa,SenthilLevin} and their associated topologically protected edge states.

%%%%%%%%%%%%%%%%%%%%%%%%%%%
\section{Magnon transport in insulating bulk magnets}
\label{sec:3dim}
%%%%%%%%%%%%%%%%%%%%%%%%%%%

In 2010, Kajiwara $\textit{et al}$.\cite{spinwave} experimentally demonstrated that it is possible to electrically create and read-out a spin-wave spin current in the magnetic insulator Y$_3$Fe$_5$O$_{12}$ (YIG) by using both inverse \cite{ISHE1} and spin Hall effects in Pt/YIG/Pt system.
The weak spin damping of YIG enables the spin-wave spin current to carry spin-information over distances of several millimeters, much further than what is typically possible when using spin-polarized conduction electrons in metals.

%%%%%%%%%%%%%%%%%%%%%%%%%%%\check{} 
\subsection{Fundamentals of Magnon Physics}
\label{subsec:3dim}
%%%%%%%%%%%%%%%%%%%%%%%%%%%
%[Dear Daniel] One of the aims of this review will to share the understanding with experiments, and this section will useful for it. Also for nonspecialist, this will be instructive.

Thus,  spin-waves (i.e., magnons \cite{BlochMagnon,HP}) play an essential role for magnetic currents in three-dimensional insulating magnets. 
Note that these systems are completely free from the restriction by the Mermin-Wagner-Hohenberg theorem \cite{MW,Hohenberg,DLFabioLeggett}; the spontaneous symmetry breaking in three dimensions is possible even in the thermodynamic limit. To reveal the transport properties of such low-energy magnetic modes is the main aim of this section.

To this end, we here (Sec. \ref{subsec:3dim}) quickly review fundamentals \cite{altland,DanielMagnetismBook,mahan,peskin,coleman,HalperinSW,Glauber,mehta,CL_text,PColeman} of spin waves  
and summarize the properties inherent to Bose particles ({\it{e.g.}}, magnons).

\begin{figure}[h]
\begin{center}
\includegraphics[width=5.5cm,clip]{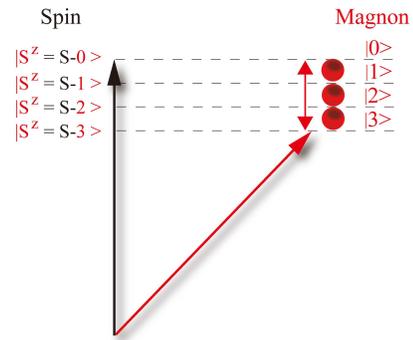}
\caption{(Color online)
Schematic representation of the correspondence between spin and magnon degrees of freedom in the Holstein-Primakoff transformation.
The magnitude of the deviation or the excitation from the ferromagnetically ordered state $ \mid  S^z = S \rangle  $ at zero temperature is represented by the number of magnons.
%and the magnetization $\langle S^z  \rangle$ is characterized by discrete values in the magnon picture.
Magnons are absent in the spin polarized state $ \mid  S^z   = S \rangle$.
\label{fig:MagnonHP} }
\end{center}
\end{figure}

%%%%%%%%%%%%%%%%%%%%%%%%%%%
\subsubsection{Holstein-Primakoff Transformation}
\label{subsubsec:HPtr}
%%%%%%%%%%%%%%%%%%%%%%%%%%%
%{\color{red} This is a very special case of a 'uniform magnon', and not really a spin wave which usually depends on a wave vector k, i.e. $a_k$ or in real space $a_i$. I think we should say this here.}
%I agree; just for simplicity, we here focus on a single mode, the zero mode k=0. Therefore I have remarked it. The purpose of this section is simply to clarify the difference from the coherent state below. Therefore I think it is enough.  

As shown in Fig. \ref{fig:spinwave}, the low-energy physics is characterized by the collective spin mode (i.e., a gapless magnetic excitation), the spin wave, and the quantum description may be identified with magnons (Fig. \ref{fig:MagnonHP}).  Assuming magnetically ordered states at zero temperature, spin degrees of freedom can be mapped into the magnon ones by the Holstein-Primakoff transformation \cite{BlochMagnon,HP}; 
focusing on the zero-mode (i.e., a single mode) just for simplicity, it becomes
%Kouki: Zero mode provides uniform spin mode, it may be identified with uniform spin-wave.
\begin{subequations}
\begin{eqnarray}
 S^+ &=& \sqrt{2S}\Big(1- \frac{a^{\dagger} a}{2S}\Big)^{1/2} a = (S^-)^{\dagger},   \label{eqn:HPdef}   \\
S^z &=& S - a^{\dagger} a,  \label{eqn:MagnonNumber}
 \end{eqnarray}
\end{subequations}
where $S$ is the spin length ($S  =  {\mathbb{N}}_+/2 \gg  1$ with ${\mathbb{N}}_+ \equiv  \{1, 2, \cdot \cdot \cdot    \}$) and the magnon annihilation (creation) operator $a^{(\dagger)}$ satisfies the bosonic commutation relation
 \begin{eqnarray}
[a, a^{\dagger}] = 1.
\end{eqnarray}
The number operator is given by $a^{\dagger} a$ and the eigenstate of the number operator, $  \mid  n  \rangle $ with $n  \in {\mathbb{N}}_0$ and ${\mathbb{N}}_0 \equiv  \{0, 1, 2, \cdot \cdot \cdot    \}$, satisfies
 \begin{subequations}
\begin{eqnarray}
 a \mid  0  \rangle &=& 0,  \\
 a^{\dagger }a  \mid  n  \rangle &=&  n  \mid  n  \rangle,  \label{eqn:eigenstate}   \\
a \mid  n  \rangle  &=&  \sqrt{n} \mid  n-1  \rangle,  \label{eqn:eigenstateminus}  \\
 a^{\dagger }\mid  n  \rangle &=&   \sqrt{n+1 }  \mid  n+1  \rangle,  \label{eqn:eigenstateplus} \\
  \langle n \mid    a \mid  n  \rangle  &=& 0.  \label{eqn:ODLROnoncondesed}
 \end{eqnarray}
\end{subequations}
Note that it is not the eigenstate of the operators $ a^{(\dagger )}$ [Eqs. (\ref{eqn:eigenstateminus}) and (\ref{eqn:eigenstateplus})].
\subsubsection{Magnonic Coherent State}
\label{subsubsec:magnoncoherent}
%%%%%%%%%%%%%%%%%%%%%%%%%%%

The way of description changes when magnons are in condensation.
Bosons are free from the Pauli principle and such a bosonic nature enables magnons to condensate\cite{BECnoteLeggett}.
Once magnons are in a condensate, they form a macroscopic coherent state and can be identified with a semiclassical object (see also Sec. \ref{subsec:outlook}); the magnon {\it{coherent state}} $\mid  \lambda  \rangle$ can be characterized by the eigenstate of the annihilation 
operator  $ a $ as
%{\color{red} Remove hats over operators, not used before!}
 \begin{subequations}
 \begin{eqnarray}
     a \mid  \lambda  \rangle &=& \lambda  \mid  \lambda  \rangle,   \\
            \lambda  &  \in &   {\mathbb{C}},
  \label{eqn:coherentstate}
\end{eqnarray}
 \end{subequations}
which means
 \begin{eqnarray}
     \langle \lambda   \mid    a   \mid    \lambda   \rangle \not= 0.
\end{eqnarray}
This is in sharp contrast to Eq. (\ref{eqn:ODLROnoncondesed}).
Thus, the expectation value of the operator $ a  $ works \cite{bunkov} as a macroscopic condensate order parameter and characterizes magnon condensation.
Using it, we classify magnon states and reveal the resulting transport properties in Sec. \ref{subsec:Josephson}.
As to the stability of the coherent state $\mid  \lambda  \rangle$ for the time evolution and the effects of magnon-magnon interactions on the coherent state, 
see Refs. [\onlinecite{Glauber,mehta,CL_text,PColeman}].
%Dear Daniel, Glauber et al. have clarified the condition for the stable coherent state as follows; $\dot{a} $ = F(a, t). If  the function F includes $a^{\dagger }$ as F(a, a^{\dagger }, t), the coherent state becomes unstable in a finite time and is destroyed.
%the properties of the coherent state  and the stability. {\color{red} stability of what?}

%\begin{figure}[h]
%\begin{center}
%\includegraphics[width=8.8cm,clip]{SSB.eps}
%\caption{(Color online)
%Schematic representation of a spontaneous symmetry breaking in the Nambu-Goldstone model $V_{\rm{NG}}^{\rm{cla.}} (\psi ) $ [Eq. (\ref{eqn:NGmodelCla})]
%for $J_0 = 10$ with (a) $\mu _0  = 1 $ and (b) $\mu _0  = -1 $. 
%Under the $U(1)$-symmetric model, the minimum value of the quantity $\psi $ that is not invariant under the global $U(1)$-symmetry transformation becomes nonzero in %the Mexican hat potential (b) when $\mu _0  < 0$. The state becomes unstable when $J_0 < 0$.
%\label{fig:SSB} }
%\end{center}
%\end{figure}

%%%%%%%%%%%%%%%%%%%%%%%%%%%
\subsection{Wiedemann-Franz Law for Magnon Transport}
\label{subsec:WF}
%%%%%%%%%%%%%%%%%%%%%%%%%%%

%%%%%%%%%%%%%%%%%%%%%%%%%%%
\subsubsection{Wiedemann-Franz Law}
\label{subsubsec:WFcharge}
%%%%%%%%%%%%%%%%%%%%%%%%%%%

In 1853, Wiedemann and Franz \cite{WFgermany} experimentally established that at sufficiently low temperatures,
the ratio between the electric and thermal conductances $\sigma$ and $K_e$ of free electrons, respectively, 
approximately reduces to the same value for different metals.
In 1872, Lorenz  discovered that the ratio becomes proportional to temperatures and now \cite{AMermin,kittel},
the Wiedemann-Franz (WF) law is summarized as 
\begin{eqnarray}
  \frac{K_e}{\sigma}  \stackrel{\rightarrow }{=}   {\cal{L}}  T,
\label{eqn:WFel} 
\end{eqnarray}
where the Lorenz number is defined by
\begin{eqnarray}
  {\cal{L}}   \equiv    \frac{{\pi}^2}{3} \Big(\frac{k_{\rm{B}}}{e}\Big)^2.   
\label{eqn:Lorenz} 
\end{eqnarray}
This is a fundamental hallmark that characterizes the universal thermoelectric properties of electron transport;
the Lorenz number is independent of any material parameters. 
Note that since electrons are fermions (Table \ref{tab:correspondence}), 
the WF law for electron transport can be characterized \cite{AMermin} only by diagonal transport coefficients $L_e^{11}$ and $L_e^{22}$ 
at low temperatures $ T  \ll  \epsilon _{\rm{F}}/k_{\rm{B}}$ ($\sim 10^4$ K for a typical metal such as gold or copper), {\it i.e.},
\begin{eqnarray}
  \frac{K_e}{\sigma}  \approx  \frac{L_e^{22} + {\cal{O}}\big((k_{\rm{B}}T/\epsilon _{\rm{F}})^2\big)}{L_e^{11}},
\label{eqn:WFel2} 
\end{eqnarray}
where $\epsilon _{\rm{F}}$ is the Fermi energy and $L_e^{ij}$ ($i, j =1 ,2$) denotes the Onsager coefficient of electron transport that characterizes the response to driving forces ({\it{e.g.}}, electric field and gradients of chemical potential and temperature; see Ref. [\onlinecite{AMermin}] for details). 
%{\color{red} no brackets (?): Ref. \onlinecite{AMermin}, fix everywhere!}

\begin{figure}[h]
\begin{center}
\includegraphics[width=8.5cm,clip]{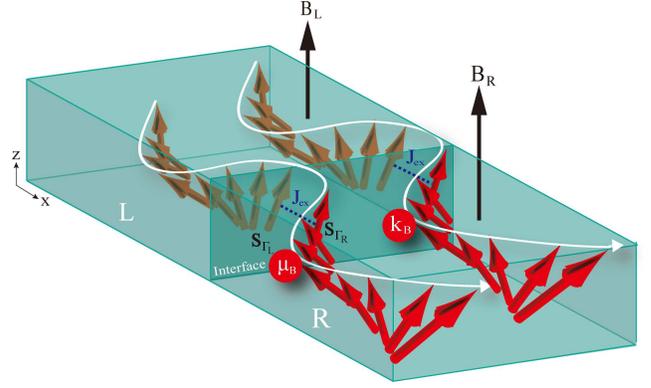}
\caption{(Color online)
Schematic representation of a three-dimensional ferromagnetic insulating junction.
%Individual circles represent noncondensed magnons and each cloud of circles is a quasiequilibrium magnon condensate.
The time reversal symmetry is broken by the ferromagnetic order and the magnetic field along the $z$-axis denoted by $ B_{\rm{L(R)}}$. 
The boundary spins ${\bf S}_{\Gamma_{\rm{L(R)}}}$ are relevant to the magnon transport
and they are weakly exchange-coupled with the strength $J_\textrm{ex}$.
Reprinted with permission from Ref. [\onlinecite{magnonWF}].
\label{fig:WF2} }
\end{center}
\end{figure}

%%%%%%%%%%%%%%%
%Start magnontransport
%%%%%%%%%%%%%%%
On the other hand (Table \ref{tab:correspondence}), magnons are bosons, without  Pauli principle and thus without Fermi surface which defines the Fermi energy.
It is well-known that the quantum-statistical properties of bosons and fermions show in general an entirely different dependence on system parameters, most notably on temperature;
as an example, at low temperatures below Fermi and Debye temperatures, the specific heat $C_V$ at constant volume of metals is given by \cite{kittel}
$  C_V =  {\cal{C}}_{\rm{el}} T +  {\cal{C}}_{\rm{ph}} T^3 $,
where electrons (i.e., fermions) contribute to ${\cal{C}}_{\rm{el}} T$, while phonons (i.e.,  bosons) to ${\cal{C}}_{\rm{ph}} T^3$ 
with a constant ${\cal{C}}_{\rm{el(ph)}}$ that depends on material properties.
Taking this into account, a fundamental question arises: can such a linear-in-$T$ behavior [Eq. (\ref{eqn:WFel})] also arise for magnons?
The answer is positive as we discuss next.
%To provide the answer and reveal thermomagnetic properties  of magnon transport is the main purpose.

%%%%%%%%%%%%%%%%%%%%%%%%%%%
\subsubsection{Onsager Matrix of Magnon}
\label{subsubsec:Onsager}
%%%%%%%%%%%%%%%%%%%%%%%%%%%

To this end, we \cite{magnonWF} considered a magnetic junction formed by two ferromagnetic insulators (FI), as illustrated in Fig. \ref{fig:WF2},
and investigated the transport of magnons. 
Focusing on thermally-induced magnons (i.e., noncondensed magnons; see Table \ref{tab:magnon}), we found universal thermomagnetic properties of magnon transport.

%The temperature of the left (right) FI is $T_{l(r)}$ and the cross-section area of the junction interface is $\cal{A}$.
Due to a finite overlap of the wave functions of the boundary spins, denoted as  ${\bf S}_{\Gamma_{\rm{L}}}$ and ${\bf S}_{\Gamma_{\rm{R}}}$ in the left and right FI, respectively (see Fig. \ref{fig:WF2}), there exists in general a finite exchange interaction described by the Hamiltonian 
\begin{eqnarray}
{\cal{H}}_{\rm{ex}}  = -J_{\rm{ex}} \sum_{\langle \Gamma_{{\rm{L}}} \Gamma_{{\rm{R}}} \rangle} 
{\bf S}_{\Gamma_{{\rm{L}}}} \cdot {\bf S}_{\Gamma_{{\rm{R}}}},
\label{eqn:Hex} 
\end{eqnarray}
where $ J_{\rm{ex}} > 0$ denotes the magnitude of the exchange interaction, weakly coupling the two FIs.
Assuming large spins $S \gg  1$ and using the linearized Holstein-Primakoff expansion \cite{HP} $ S^+  \approx  \sqrt{2S} a$,  the tunneling Hamiltonian can be written by  magnon degrees of freedom
\begin{eqnarray}
 {\cal{H}}_{\rm{ex}}^{\rm{tun}} =  - J_{{\rm{ex}}} S \sum_{{\mathbf{k}}_{\perp }}  \sum_{k_x, k_x^{\prime}}
a_{\Gamma _{{\rm{L}}},  {\mathbf{k}}}  a^{\dagger }_{\Gamma _{{\rm{R}}}, {\mathbf{k}}^{\prime}} + {\rm{H. c.}},
\label{eqn:nonconservedScattering} 
\end{eqnarray}
where ${\mathbf{k}}=(k_x, k_y, k_z)$, ${\mathbf{k}}^{\prime}=(k_x^{\prime}, k_y, k_z)$, ${\mathbf{k}}_{\perp }=(0, k_y, k_z)  $,
and 
the bosonic operator $a_{\Gamma_{{\rm{R/L}}}}^{\dagger}$ ($a_{\Gamma_{{\rm{R/L}}}}$) creates (annihilates) a boundary magnon at the right/left FI. 
The $k_x$-momentum of magnons is not conserved at the sharp junction interface, whereas the perpendicular momentum ${\mathbf{k}}_{\perp }$ is conserved. 
We microscopically confirmed that the thermomagnetic properties of magnon transport in the junction qualitatively remain valid even when 
$  k_x  =  k_x^{\prime}$ in a model; the properties are robust against the microscopic details of the junction.

The tunneling Hamiltonian  ${\cal{H}}_{\rm{ex}}^{\rm{tun}} $ thus gives the time-evolution of the magnon number operators in both FIs 
and generates the magnetic and heat currents.
The temperature of the left (right) FI is $T_{{\rm{L(R)}}}$ and  the magnetic field for magnons in the left (right) FI along the $z$-axis is  $B_{{\rm{L(R)}}} $
with redefining $ B_{\rm{L}}\equiv B$ and $T_{\rm{L}}\equiv T$ for convenience.
The magnetic field and temperature differences are then defined by 
$\Delta B \equiv  B_{{\rm{L}}} -  B_{{\rm{R}}}$ and $   \Delta T \equiv  T_{{\rm{L}}} -  T_{{\rm{R}}}$, respectively,
and they generate the magnetic and heat (i.e., energy) currents  \cite{AMermin,mahan}  ${\cal{I}}_{\rm{m}} $ and ${\cal{I}}_{Q}$
\begin{subequations}
\begin{eqnarray}
{\cal{I}}_{\rm{m}}& =& - i  (J_{\rm{ex}}S/\hbar ) 
\sum_{\mathbf{k},{\mathbf{k}}^{\prime}} 
g \mu _{\rm{B}}  a_{\Gamma_{{\rm{L}}},{\mathbf{k}}}  a^{\dagger }_{\Gamma_{{\rm{R}}},{\mathbf{k}}^{\prime}} + {\rm{H. c.}}, 
\label{eqn:Im}  \\
%%%%%%%%%%%%%%%%%%%%%%%%%%%%%%%%
{\cal{I}}_{Q} &= &- i  (J_{\rm{ex}}S/\hbar ) 
\sum_{\mathbf{k},{\mathbf{k}}^{\prime}} 
\omega _{{{ {k}}}}^{\rm{L}}  a_{\Gamma_{{\rm{L}}},{\mathbf{k}}}  a^{\dagger }_{\Gamma_{{\rm{R}}},{\mathbf{k}}^{\prime}} + {\rm{H. c.}},
\label{eqn:IQ} 
\end{eqnarray}
\end{subequations}
%%%%%%%%%%%%%%%%%%%%%%%
where $\omega _{{{ {k}}}}^{\rm{L}}$ is the magnon dispersion relation in the left FI (see Ref. [\onlinecite{magnonWF}] for the detailed expression).
The currents flow from the right FI to the left one when the sign of the currents is positive.
Within the linear response regime, each Onsager coefficient $L^{ij}$ ($i, j = 1, 2$) is defined by
\begin{eqnarray}
%%%%%%%%%%%%%%%%%%%%%%%%%%%%%
\begin{pmatrix}
\langle {\cal{I}}_{\rm{m}} \rangle  \\  \langle {\cal{I}}_{Q} \rangle
\end{pmatrix}
%%%%%%%%%%%%%%%%%%%%%%%%%%%%%
=
%%%%%%%%%%%%%%%%%%%%%%%%%%%%%
\begin{pmatrix}
L^{11} & L^{12} \\ L^{21} & L^{22}
\end{pmatrix}
%%%%%%%%%%%%%%%%%%%%%%%%%%%%%
%%%%%%%%%%%%%%%%%%%%%%%%%%%%%
\begin{pmatrix}
    \Delta B  \\    -  \Delta T
\end{pmatrix}
,
%%%%%%%%%%%%%%%%%%%%%%%%%%%%%
\label{eqn:LinearResponseIJ}
\end{eqnarray}
%%%%%%%%%%%%%%%%%%%%%%%%%%%%%%
and it is evaluated by a straightforward perturbative calculation up to ${\cal{O}} (J_{\rm{ex}}^2)$ in $J_{\rm{ex}}$  based on the Schwinger-Keldysh formalism\cite{Schwinger,Schwinger2,Keldysh,rammer,haug,kamenev,tatara,kita,new,LuttingerNOTE}. 
See Ref. [\onlinecite{magnonWF}] for detailed expressions of the Onsager coefficients, which depend on material parameters.

Both currents arise from terms of order $ {\cal{O}} (J_{\rm{ex}}^2)$.
Therefore, even when an electric field is applied to the interface, the resulting A-C phase cannot play any significant  role in the transport of such noncondensed magnons. 
Moreover, even when a magnetic field difference $ \Delta B \ne 0$ is generated, the noncondensed magnon current becomes essentially a dc one (Table \ref{tab:magnon}). 
These are in sharp contrast to the condensed magnon current \cite{KKPD}, 
which arises from the $ {\cal{O}}(J_{\rm{ex}})$-term and can become an ac current (see Sec. \ref{subsec:Josephson}).

\begin{figure}[h]
\begin{center}
\includegraphics[width=6cm,clip]{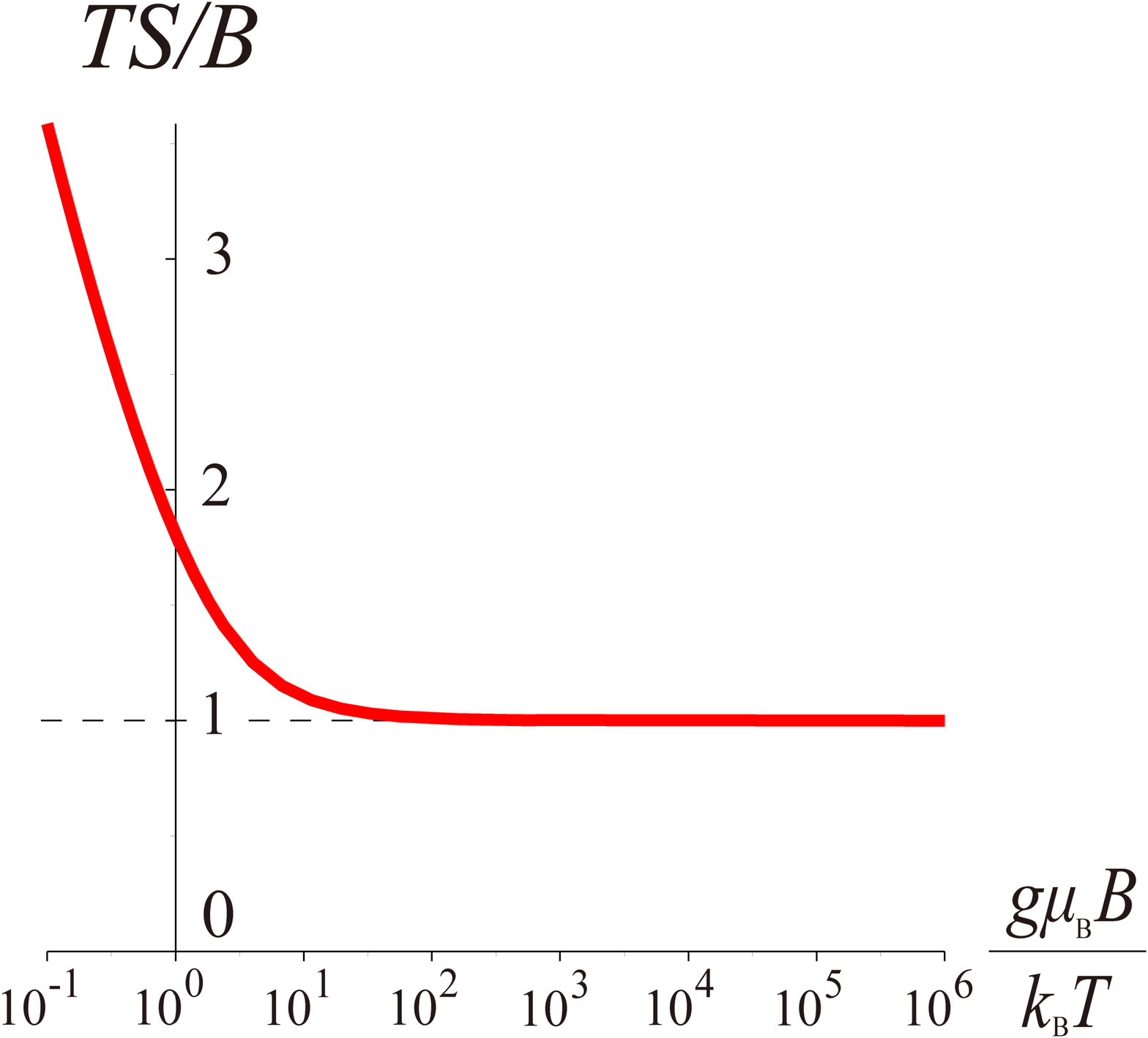}
\caption{(Color online)
Plots of magnon Seebeck coefficient $\cal{S}$  as function of $ g \mu _{\rm{B}}B/(k_{\rm{B}} T)$.
At low temperatures $ g \mu _{\rm{B}}B/(k_{\rm{B}} T)  = {\cal{O}}(10^2)$, the rescaled coefficient $ T {\cal{S}}/B $ approaches unity asymptotically and realizes the universal relation.
Reprinted with permission from Ref. [\onlinecite{magnonWF}].
\label{fig:WFSeebeckmagnon} }
\end{center}
\end{figure}

%%%%%%%%%%%%%%%%%%%%%%%%%%%%%%%
\subsubsection{Magnon Seebeck and Peltier Coefficients}
\label{subsubsec:SeebeckPeltier}
%%%%%%%%%%%%%%%%%%%%%%%%%%%%%%%

The time reversal symmetry is broken by the ferromagnetic order and the magnetic field,
and therefore the Onsager reciprocal relation \cite{Onsager2,Casimir,OnsagerRelation} could in principle be violated.
Still, we microscopically found that the relation is satisfied (see Sec. \ref{subsubsec:remarkOnsager} for details)
\begin{eqnarray}
 L^{21}=   T\cdot  L^{12}.
  \label{eqn:OnsagerRelation}
\end{eqnarray}
This provides the Thomson relation (i.e., the Kelvin-Onsager relation \cite{spincal}) 
\begin{eqnarray}
  \Pi  = T {\cal{S}},
  \label{eqn:Thomson}
\end{eqnarray}
where 
${\cal{S}} \equiv  L^{12}/L^{11}$ is the magnon Seebeck coefficient 
and $\Pi  \equiv  L^{21}/L^{11}$ is the  magnon Peltier coefficient.
%%%%%%%%%%%%%%%%%%%%%%%%%%
At low temperatures, $   \hbar /(2\tau )  \ll    k_{\rm{B}} T \ll  g \mu _{\rm{B}} B$ with a phenomenologically introduced lifetime $\tau $ of magnons mainly due to nonmagnetic impurity scatterings,  the coefficients reduce to (Fig. \ref{fig:WFSeebeckmagnon})
\begin{eqnarray}
 {\cal{S}} \stackrel{\rightarrow }{=} \frac{B}{T}, \,\,\,\,\,\,\,\,\,\,\,\,\,
 \Pi  \stackrel{\rightarrow }{=} B.
\label{eqn:Peltier}  
\end{eqnarray}
The magnon Seebeck and Peltier coefficients thus become {\it universal} at low temperatures; 
including the $g$-factor, they are completely independent of any material parameters and are solely determined by the applied magnetic field and temperature.

%%%%%%%%%%%%%
\begin{table*}
\caption{
\label{tab:correspondence}
Thermomagnetics of magnon transport.
The time reversal symmetry is broken by the ferromagnetic order and the magnetic field, but the Onsager reciprocal relation is still satisfied.
The Onsager relation ensures the Thomson relation, and vice versa.
The magnon Seebeck and Peltier coefficients become universal at low temperatures $   \hbar /(2\tau )  \ll    k_{\rm{B}} T \ll  g \mu _{\rm{B}} B $ in the sense that  
including the $g$-factor, they are completely independent of any material parameters and are solely determined by the applied magnetic field and temperature.
%%%%%%%%%%%%%%%%%%%%%%%%%%%%
The WF law for electron transport is characterized \cite{AMermin,kittel} only by diagonal elements $L_e^{11}$ and $L_e^{22}$ 
due to the relatively large Fermi energy ${\cal{O}}\big((k_{\rm{B}}T/\epsilon _{\rm{F}})^2\big)$,
while off-diagonal elements $L^{21}$ and $L^{12}$ are essential to the thermal conductance and the WF law for magnon transport since magnons are bosons.
Still, the $T$-linear behavior holds in the same way for magnons despite the difference of quantum-statistical properties;
the role of the charge $e$ is played by $g \mu _{\rm{B}}$ and the magnetic Lorenz number is independent of any material parameters except the $g$-factor which is material specific. 
}
%%%%%%%%%%%%
\begin{ruledtabular}
\begin{tabular}{cccccc}
%%%%%%%%%%%%
   &  Electron &  Magnon   \\ \hline
 Statistics  &Fermi-Dirac  & Bose-Einstein  \\
Electric and magnetic conductance   &   $\sigma  \equiv L_e^{11} $, &   $ G \equiv L^{11} $  \\
Thermal conductance   &   $ K_e  \approx  L_e^{22} + {\cal{O}}\big((k_{\rm{B}}T/\epsilon _{\rm{F}})^2\big) $, 
&   $ K \equiv  L^{22} -  L^{21} L^{12}/L^{11}$  \\
WF law (low temperature)   &   $ K_e/\sigma  \approx  [L_e^{22} + {\cal{O}}\big((k_{\rm{B}}T/\epsilon _{\rm{F}})^2\big)]/L_e^{11} 
\stackrel{\rightarrow }{=} {\cal{L}}  T$, 
&   $ K/G \equiv  (L^{22} -  L^{21} L^{12}/L^{11})/L^{11} \stackrel{\rightarrow }{=}   {\cal{L}}_{\rm{m}}  T$  \\
Lorenz number & ${\cal{L}} =\frac{{\pi}^2}{3} \Big(\frac{k_{\rm{B}}}{e}\Big)^2 $    
& ${\cal{L}}_{\rm{m}} =   \Big(\frac{k_{\rm{B}}}{g \mu _{\rm{B}}}\Big)^2$  \\
Seebeck (${\cal{S}}$) and Peltier ($\Pi$) coefficients &  ${\cal{S}} \equiv  L_e^{12}/L_e^{11}$,  $\Pi  \equiv  L_e^{21}/L_e^{11}$  
&  ${\cal{S}} \stackrel{\rightarrow }{=} \frac{B}{T}$, $\Pi  \stackrel{\rightarrow }{=} B$ \\
Onsager relation & $L_e^{21}=   T\cdot  L_e^{12}$  &  $L^{21}=   T\cdot  L^{12}$ \\
Thomson relation & $\Pi  = T {\cal{S}}$ & $\Pi  = T {\cal{S}}$ \\
\end{tabular}
\end{ruledtabular}
\end{table*}
%%%%%%%%%%

%%%%%%%%%%%%%%%%%%%%%%%%%%%%%%%%%
\subsubsection{Thermal Conductance for Magnon Transport}
\label{subsubsec:ThermalConductance}
%%%%%%%%%%%%%%%%%%%%%%%%%%%%%%%%%

The number of magnons increases when temperature becomes higher and magnetic fields weaker.
%such a condition is comfortable for magnons.
Under the condition $ \Delta  T =0$,
the magnetic conductance $G$ is defined by
\begin{eqnarray}
\langle {\cal{I}}_{\rm{m}} \rangle \equiv   G \cdot  \Delta  B,
\end{eqnarray}
which results in [Eq. (\ref{eqn:LinearResponseIJ})]
\begin{eqnarray}
G =  L^{11}. 
\end{eqnarray}
The thermal conductance $K$ is defined by \cite{AMermin,mahan}
\begin{eqnarray}
\langle {\cal{I}}_{Q} \rangle \equiv  - K \cdot  \Delta  T      \     \    {\rm{with}}   \    \     \langle {\cal{I}}_{\rm{m}} \rangle = 0,
\end{eqnarray}
which can be understood as follows\cite{BassoNOTE};
the applied temperature difference produces a magnon current and consequently a magnetization difference \cite{SilsbeeMagnetization,Basso,Basso2} that induces a counter current.
Thus, the system reaches a new quasi-equilibrium steady state where magnon currents no longer flow.
Since the condition for such quasi-equilibrium state $ \langle {\cal{I}}_{\rm{m}} \rangle = 0$ results in a magnetization difference $ \triangle B^{\ast } = (L^{12}/L^{11}) \Delta  T $ that gives the counter current, the thermal conductance is given by [Eq. (\ref{eqn:LinearResponseIJ})] 
\begin{eqnarray}
K = L^{22} -  \frac{L^{21} L^{12}}{L^{11}}.
\end{eqnarray}
Note that in sharp contrast to charge transport [Eq. (\ref{eqn:WFel2})],
off-diagonal elements $L^{21}$ and $L^{12}$ which correspond to the counter current play a key role in the thermal conductance for magnon transport (see Table \ref{tab:correspondence}) since magnons are bosons and thus $L^{21}$ and $L^{12}$ are not suppressed by a large Fermi energy.
The information from all elements of the Onsager matrix [Eq. (\ref{eqn:LinearResponseIJ})] is essential for the thermal conductance of bosons.

\begin{figure}[h]
\begin{center}
\includegraphics[width=6cm,clip]{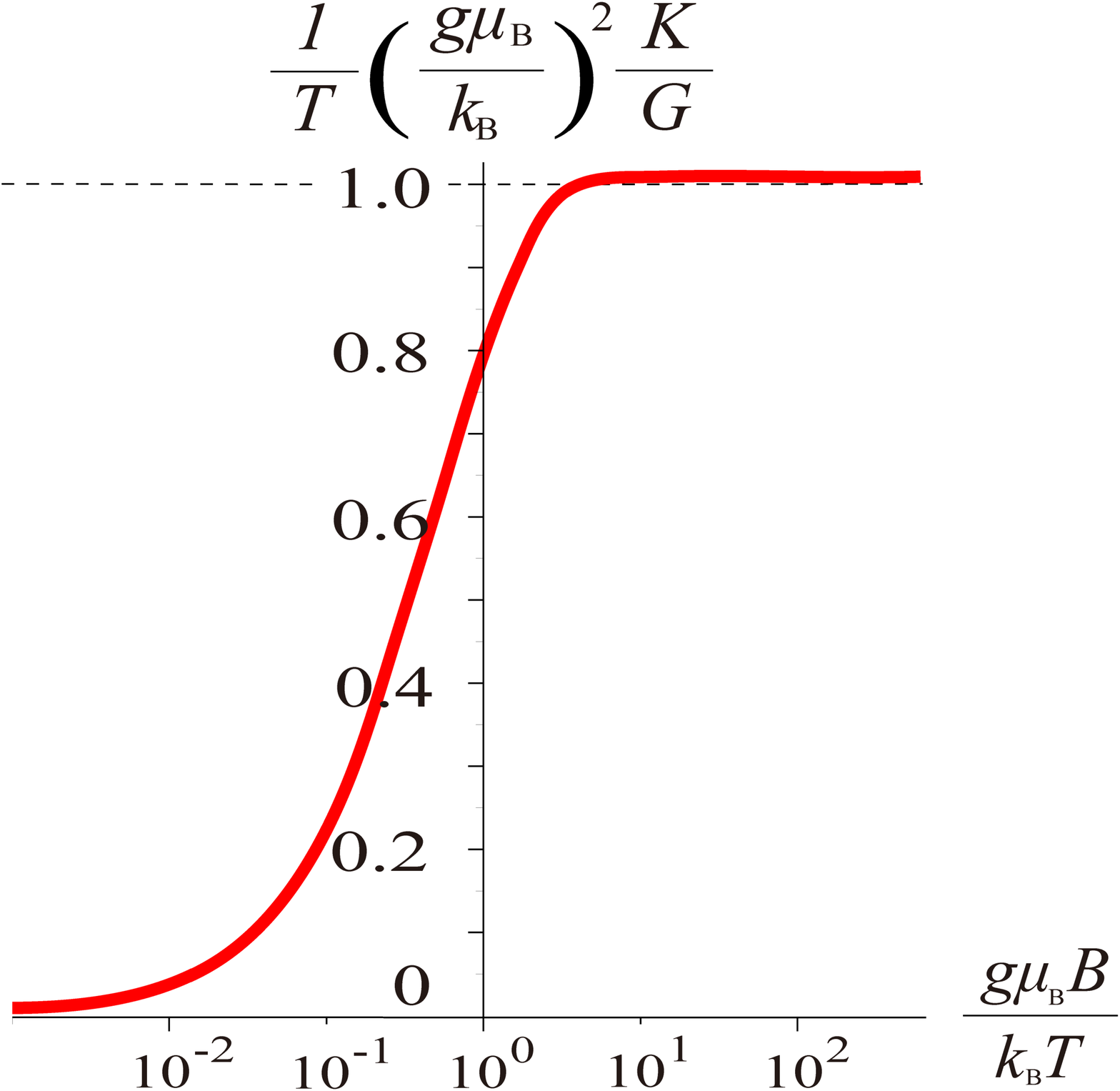}
\caption{(Color online)
Plots of the ratio $ (g \mu _{\rm{B}}/k_{\rm{B}})^2[K/(GT)] $ as function of $ g \mu _{\rm{B}}B/(k_{\rm{B}} T)$.
The ratio reaches the constant `$1$' and the WF law  for magnon transport is realized at low temperatures $ g \mu _{\rm{B}}B/(k_{\rm{B}} T)= {\cal{O}}(10)$;  
otherwise, the linear-in-$T$ behavior breaks down.
Reprinted with permission from Ref. [\onlinecite{magnonWF}].
\label{fig:WFSeebeckmagnon2} }
\end{center}
\end{figure}

Microscopic calculation shows that at low temperatures $ \hbar /(2\tau )  \ll      k_{\rm{B}} T \ll  g \mu _{\rm{B}} B$, 
the ratio  becomes  linear in temperature (Fig. \ref{fig:WFSeebeckmagnon2})
\begin{eqnarray}
 \frac{K}{G}  \stackrel{\rightarrow }{=}   \Big(\frac{k_{\rm{B}}}{g \mu _{\rm{B}}}\Big)^2    T.
 \label{eqn:WFmagnon}
\end{eqnarray}
Therefore, in analogy to charge transport in metals \cite{WFgermany,AMermin,kittel}, 
we refer to this behavior as the {\it Wiedemann-Franz law for magnon transport}.
The constant ${\cal{L}}_{\rm{m}}$ analogous to the Lorenz number is given by
\begin{eqnarray}
 {\cal{L}}_{\rm{m}}   \equiv  \Big(\frac{k_{\rm{B}}}{g \mu _{\rm{B}}}\Big)^2,    
 \label{eqn:Lorenzmagnon}
\end{eqnarray}
where the role of the charge $e$ is played by $g \mu _{\rm{B}}$.
Therefore we refer to this constant as the {\it magnetic Lorenz number}.
The magnetic Lorenz number is independent of any material parameters except the $g$-factor which is material specific. 
Interestingly, the $T$-linear behavior holds in the same way for magnons despite the fundamental difference between the quantum-statistical properties of bosons and fermions.
%magnons are bosonic excitations, while electrons are fermions. 
The thermomagnetic properties are summarized in Table \ref{tab:correspondence}.
%%%%%%%%%%

%I would like to stress the folloing points to appeal our significance
One might suspect that such a $T$-linear behavior can be qualitatively expected from the unit conversion.
However, without the microscopic calculations, one can not eliminate the possibility that the ratio becomes 
$K/G \propto  T [k_{\rm{B}}T/(g\mu _{\rm{B}}B)]^{m-1} \propto T^{m}  $ with $m$ some integer 
even within linear response theory [Eq. (\ref{eqn:LinearResponseIJ})]
since the driving force for magnon transport is not $B$ but $\Delta B$; consequently, each Onsager coefficient is characterized by an expansion of the dimensionless quantity $g\mu _{\rm{B}}B/(k_{\rm{B}}T)$. 
See Fig. \ref{fig:WFSeebeckmagnon2}.
Indeed, if one (wrongly) omits the off-diagonal elements $L^{12}$ and $L^{21}$, the ratio $K/G$ is not proportional to temperature and reduces to a wrong result.
%%%%%%%%%%%%%%%%%%
Thus the linear-in-$T$ behavior can be traced back to the Onsager relation between the off-diagonal elements [Eq. (\ref{eqn:OnsagerRelation})].
The magnon WF law becomes violated when the Onsager relation is broken.

%%%%%%%%%%%%%%%%%%%%%%%%%%%%%%%%%
\subsubsection{Magnon-magnon interaction}
\label{subsubsec:magmagint2}
%%%%%%%%%%%%%%%%%%%%%%%%%%%%%%%%%

We microscopically found that the anisotropy induced magnon-magnon interaction $J_{\rm{m}}$ provides a `nonlinearity' in terms of the perturbative terms and violates the Onsager relation (see Ref. [\onlinecite{magnonWF}] for details);
it gives contributions $\delta L^{11}  $ and $ \delta L^{21}$, and the Onsager matrix in Eq. (\ref{eqn:LinearResponseIJ}) is replaced by
\begin{equation}
%%%%%%%%%%%%%%%%%%%%%%%%%%%%%
\begin{pmatrix}
  L^{11}   &   L^{12}  \\ 
  L^{21}  &  L^{22} 
\end{pmatrix}
\mapsto 
%%%%%%%%%%%%%%%%%%%%%%%%%%%%%
\begin{pmatrix}
  L^{11} + \delta L^{11}  &   L^{12}  \\ 
  L^{21} + \delta L^{21} &  L^{22} 
\end{pmatrix},
%%%%%%%%%%%%%%%%%%%%%%%%%%%%%
\label{eqn:violate}
\end{equation}
where $\delta  L^{11}  =  {\cal{O}}(J_{\rm{ex}}^2J_{\rm{m}})$ and $\delta  L^{21}  =  {\cal{O}}(J_{\rm{ex}}^2J_{\rm{m}})$, while $  L^{11}  =  {\cal{O}}(J_{\rm{ex}}^2J_{\rm{m}}^0)$ and $  L^{21}  =  {\cal{O}}(J_{\rm{ex}}^2J_{\rm{m}}^0)$.
This can be understood as follows (see Appendix D in Ref. [\onlinecite{magnonWF}] for details);
the magnon-magnon interaction acts as an effective magnetic field and in an external magnetic field difference $\Delta B $, the total magnetic field difference $ \Delta B_{\rm{tot}} $ becomes $ \Delta B_{\rm{tot}} = (1+b_{\rm{m}})\Delta B  $ with $ b_{\rm{m}}={\cal{O}}(J_{\rm{m}})$.
The term $ b_{\rm{m}}$ gives $\delta  L^{11}={\cal{O}}(J_{\rm{m}})$ and $\delta  L^{21}={\cal{O}}(J_{\rm{m}})$,
and the magnitude of the effective magnetic field difference can be estimated by  $  b_{\rm{m}} \sim \delta L^{21}/L^{21}$.
The Onsager relation is thus violated due to the nonlinearity caused by the anisotropy induced magnon-magnon interaction.
%%%%%%%%%%%%%%%%%%%%%%%%%%%%
However, at low temperatures, where the WF law and the universality of Seebeck and Peltier coefficients hold ({\it{e.g.}}, $T = 0.7$K and $ B  = 5$T), these deviations become negligible  $|\delta L^{ij}/L^{ij}|\ll 1$ and consequently, the Onsager relation and the WF law approximately hold.
At such low temperatures $T\sim 10^{-1}$K, phonon contributions \cite{WFphononAF} are negligible \cite{adachiphonon}.

%%%%%%%%%%%%%%%%%%%%%%%%%%%%%%%%%
\subsubsection{Universality}
\label{subsubsec:universality}
%%%%%%%%%%%%%%%%%%%%%%%%%%%%%%%%%

So far we have assumed bulk FIs where magnetic dipole-dipole interactions are negligible \cite{tupitsyn}. 
Such dipolar effects, however, become important in thin films, resulting in a modified dispersion for magnons 
where the lowest energy mode becomes $k =  k_{\rm{m}}  \sim  10^4 $/cm for YIG thin films \cite{demokritov,ultrahot,SergaBEC};
the magnitude of $k_{\rm{m}}$ depends on the width of YIG thin films.
We microscopically confirmed \cite{magnonWF} that the WF law (i.e., the $T$-linear behavior) still remains valid in this case too, underlining the universality of this law.
See Ref. [\onlinecite{magnonWF}] for details.

We mention that in sharp contrast to the three-dimensional insulating ferromagnets, 
quasiparticle pictures ({\it{e.g.,}} magnons) become broken down due to strong quantum fluctuations in the one-dimensional integrable spin-$1/2$ $XXZ$ chain.
Still, the linear-in-$T$ behavior is satisfied \cite{WFchain} at low temperatures; it does not depend on the dimensionality.
Therefore the $T$-linear behavior may be identified with the universal thermomagnetic properties of spin transport.
See Refs. [\onlinecite{WFnFL,WFnFL2,WFnFL3,WFnFL4,WFnFL5}] for the progress\cite{FQHEnote} on the WF law for non-Fermi liquids.

%%%%%%%%%%%%%%%%%%%%%%%%%%%%%%%%%
\subsubsection{Remark: Onsager-Thomson Relation}
\label{subsubsec:remarkOnsager}
%%%%%%%%%%%%%%%%%%%%%%%%%%%%%%%%%

Our microscopic calculation (see Ref. [\onlinecite{magnonWF}] for details) showed that magnetic and heat currents [Eqs. (\ref{eqn:Im}) and (\ref{eqn:IQ})] are characterized by the difference of  Bose-distribution functions $n_{\rm{B}} (\omega)$.
Expanding in powers of $\mid  \Delta T  \mid  \ll T$ and $\mid  \Delta B \mid  \ll B$ within the linear response regime,
the difference becomes [$\beta \equiv  1/(k_{\rm{B}}T)$]
\begin{eqnarray}
    n_{\rm{B}} (\omega _{\mathbf{k}}^{\rm{L}}) -  n_{\rm{B}} (\omega _{\mathbf{k}}^{\rm{R}})   \approx 
\begin{cases}
  -  \beta   \frac{g \mu_{\rm{B}} e^{\beta \omega _{\mathbf{k}}}}{(e^{\beta \omega _{\mathbf{k}}}-1)^2}  \Delta  B,  
& {\textrm{ for }}   \Delta T  = 0,
  \\ 
%%%%%%%%%%%%%%%%%%%%%%%%%%%%%%%%%%%%%%%%%%%%%%%%%
  \frac{\beta  \omega _{\mathbf{k}}}{T}  \frac{e^{\beta \omega _{\mathbf{k}}}}{(e^{\beta \omega _{\mathbf{k}}}-1)^2}  \Delta  T,
& {\textrm{ for }} \Delta B  = 0,   
    \    \label{eqn:BoseDistributionGreen2}
\end{cases}
\end{eqnarray}
which yields
\begin{eqnarray}
 \frac{[n_{\rm{B}} (\omega _{\mathbf{k}}^{\rm{L}}) -  n_{\rm{B}} (\omega _{\mathbf{k}}^{\rm{R}})] \mid _{\Delta B=0}/\Delta T}
{[n_{\rm{B}} (\omega _{\mathbf{k}}^{\rm{L}}) -  n_{\rm{B}} (\omega _{\mathbf{k}}^{\rm{R}})] \mid _{\Delta T=0}/\Delta B}
 = - \frac{\omega _{\mathbf{k}}^{\rm{L}}}{g \mu_{\rm{B}}T}.
\label{eqn:BoseDistributionGreen3}
\end{eqnarray}
Eq. (\ref{eqn:BoseDistributionGreen3}) is responsible for the Onsager-Thomson relations, given in Eqs. (\ref{eqn:OnsagerRelation}) and (\ref{eqn:Thomson}).
Thus, we found that the magnetic and heat currents are characterized by the difference of the Bose-distribution functions, and
the difference gives the linear responses [Eqs. (\ref{eqn:BoseDistributionGreen2}) and (\ref{eqn:BoseDistributionGreen3})], 
and the Onsager-Thomson relations, Eqs. (\ref{eqn:OnsagerRelation}) and (\ref{eqn:Thomson}), hold accordingly.

Such results [{\it{e.g.}}, Eq. (\ref{eqn:BoseDistributionGreen2})] might be provided or expected also by the Landauer-B$\ddot{\rm{u}}$ttiker formalism \cite{kittel,Landauertext,Lesovik,Datta}.
However, the method is applicable only to noninteracting case, while the Schwinger-Keldysh formalism \cite{Schwinger,Schwinger2,Keldysh,rammer,haug,kamenev,tatara,kita,new} enabled us to perturbatively evaluate the effect of the anisotropy induced magnon-magnon interaction.

\begin{figure}[h]
\begin{center}
\includegraphics[width=3.5cm,clip]{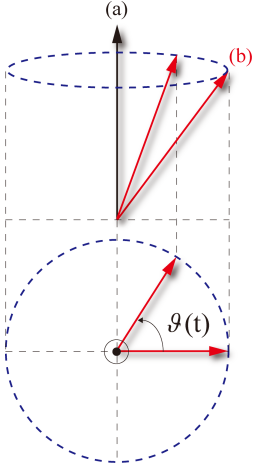}
\caption{(Color online)
(a) A spin polarized state in the ground state, where magnons are absent.
(b) A spin precession with the frequency $ d \vartheta (t)/dt\not= 0$.
In quasiequilibrium magnon condensation, all spins precess with the same finite frequency $d \vartheta  (t)/dt$ and in terms of spin variables, such quasiequilibrium magnon condensation can be identified with a macroscopic coherent spin precession, 
see Fig. \ref{fig:Incoherent}.
\label{fig:MagnonSpin} }
\end{center}
\end{figure}

%%%%%%%%%%%%%%%%%%%%%%%%%%%
\subsection{Magnon Josephson Effects}
\label{subsec:Josephson}
%%%%%%%%%%%%%%%%%%%%%%%%%%%

Next, focusing on magnons in quasiequilibrium \cite{demokritov,ultrahot,SergaBEC,KopietzBEC} condensation,
we investigate the transport properties and clarify differences from the noncondensed magnons.

%%%%%%%%%%%%%%%%%%%%%%%%%%%%%%%%%
\subsubsection{Quasiequilibrium Condensation of Excited Magnons}
\label{subsubsec:quasiequilibrium}
%%%%%%%%%%%%%%%%%%%%%%%%%%%%%%%%%

In 2006, Demokritov {\textit{et al.}} \cite{demokritov} experimentally realized {\it{quasiequilibrium}} \cite{bunkov,Yukalov,BatistaBEC,Mills} magnon condensates in YIG thin film by microwave pumping method at room temperature. 
Using Brillouin light scattering spectroscopy \cite{demokritovReport}, the relation between microwave pumping and the resulting magnon condensate was investigated in detail by Serga {\textit{et al.}} \cite{ultrahot} and Calusen {\textit{et al.}} \cite{SergaBEC}; magnetic dipole-dipole interactions become relevant in YIG thin film and the lowest energy mode of magnons  becomes nonzero  $ k_{\rm{m}} \not =0$ ({\it{e.g.}}\cite{demokritov,ultrahot,SergaBEC}, $k_{\rm{m}}  \sim  10^4 $/cm), which plays the key role on the formation of quasiequilibrium magnon condensates.
The applied microwave drives the system into a {\it{nonequilibrium}} steady state and continues to populate the zero-mode of magnons with breaking the $U(1)$-symmetry. 
After switching off the microwaves, the $U(1)$-symmetry is recovered and the number of magnons becomes conserved.
Then toward the true lowest energy state $ k_{\rm{m}} \not =0$, the system undergoes a thermalization \cite{SergaBEC} 
(or relaxation \cite{vannucchi,vannucchi2,rezende,hick,kloss,troncoso,LTP}) process and thereby reaches a metastable \cite{bunkov,Yukalov,BatistaBEC,Mills} state where the pumped magnons form a  {\it{quasiequilibrium}} magnon condensate\cite{BECnote}.
The quasiequilibrium magnon condensate is not the ground state but a metastable \cite{bunkov,Yukalov,BatistaBEC,Mills} state that corresponds to a macroscopic coherent precession \cite{bunkovalone} in terms of spin variables which can last \cite{demokritov,ultrahot,SergaBEC} for a few hundred nanoseconds.
%%%%%%%%%%%%%%%%%%%%%%
%Such realized quasiequilibrium magnon condensation is qualitatively different from the well-known thermally  {\it{equilibrium}} \cite{oshikawa,MillsExperiment} magnon condensation associated with the spontaneous $U(1)$-symmetry breaking \cite{snoke,kloss,Mills}, since the applied microwave breaks the symmetry not spontaneously but explicitly \cite{snoke,leggett}.
%The spontaneous $U(1)$-symmetry breaking and the resulting thermally equilibrium magnon condensation might be well understood \cite{kloss,hickbec} by the  Nambu-Goldstone model  with a Mexican hat potential (see Sec. \ref{subsec:3dim} and Fig. \ref{fig:SSB}), and it is realized in the thermodynamic ground state.
%%%%%%%%%%%%%%%%%%%%%%

\begin{figure}[h]
\begin{center}
\includegraphics[width=8cm,clip]{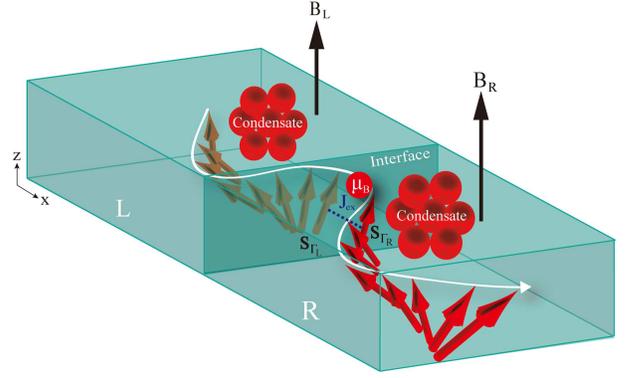}
\caption{(Color online)
Schematic representation of a magnonic Josephson junction at zero temperature.
See also Fig. \ref{fig:WF2}.
Reprinted with permission from Ref. [\onlinecite{KKPD}].
\label{fig:WF3} }
\end{center}
\end{figure}

\begin{figure}[h]
\begin{center}
\includegraphics[width=8.5cm,clip]{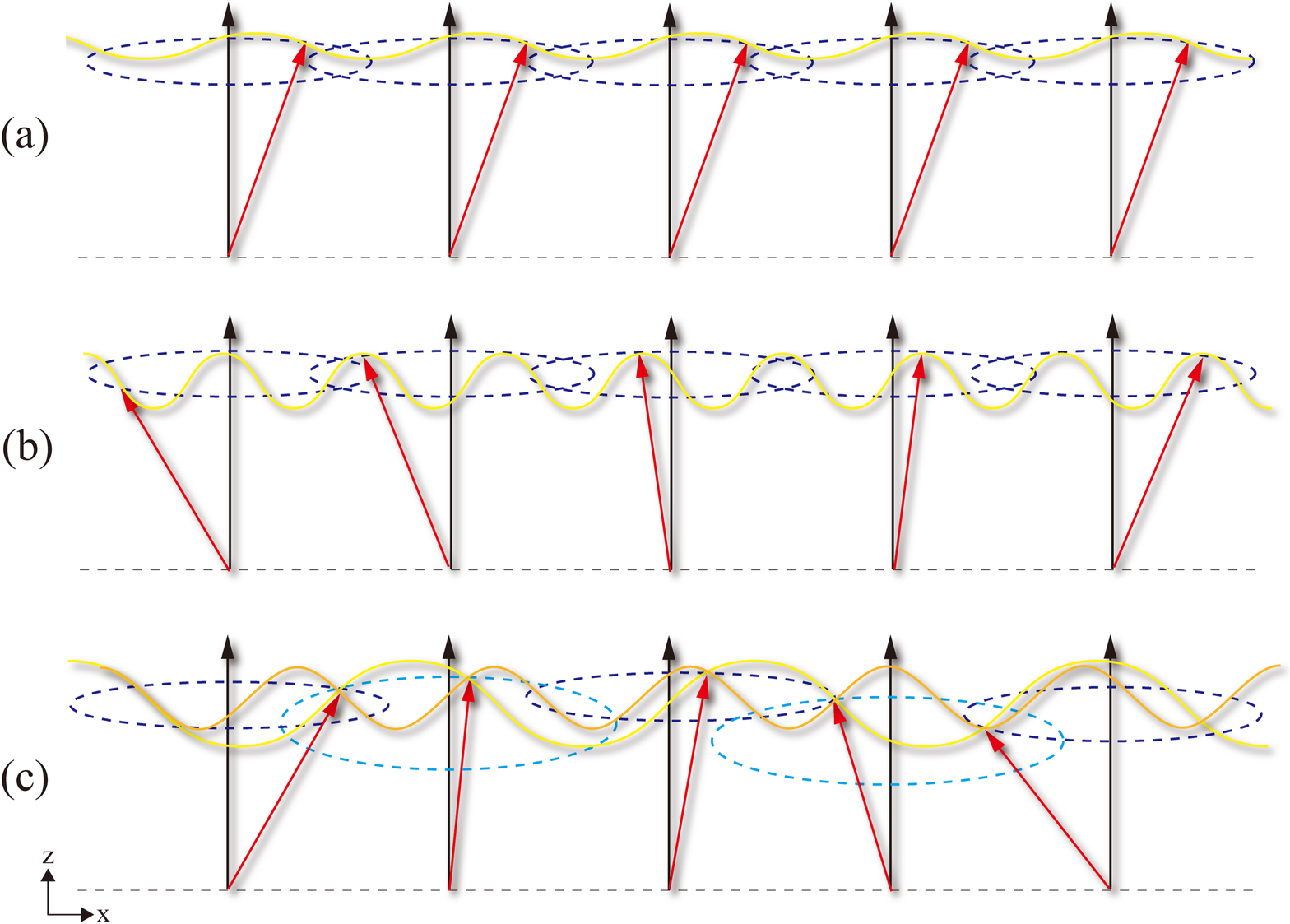}
\caption{(Color online)
Schematic representations of (a) a macroscopic coherent spin precession, (b) a macroscopic coherent spin precession in A-C effects, and (c) an incoherent spin precession.
The dashed blue lines represent a spin precession on a plane, and the frequency is different from the one in light blue.
All spins precess with the same frequency (i.e., coherently) in (a) and (b), while not in (c).
In terms of spin variables, quasiequilibrium magnon condensates \cite{demokritov,ultrahot,SergaBEC} correspond to (a) and (b).
Applying an external electric field  $   {\mathbf{E}} = E{\bf e}_y $ in (b) with a lattice constant $a$, the resulting A-C effects generate a constant phase difference $\theta _{\rm{A{\mathchar`-}C}} = [g \mu_{\rm{B}}/(\hbar c^2)] E a $ between the individual spins, but still they precess with the same frequency, which leads to a persistent current in a ring (Fig. \ref{fig:Ring}).  
\label{fig:Incoherent} }
\end{center}
\end{figure}

From a theoretical viewpoint, in sharp contrast to noncondensed magnons, such quasiequilibrium condensed magnons are characterized by a macroscopic condensate order parameter commonly called the off-diagonal long-range order (ODLRO) \cite{bunkov,PColeman,leggett}
\begin{eqnarray}
   \langle    a(t) \rangle  =  \sqrt{N_{\rm{cond.}}}  {\rm{e}}^{i \vartheta  (t)},
    \label{eqn:ODLROdef}
\end{eqnarray}
where $N_{\rm{cond.}} =\langle    a(t) \rangle ^{\ast }  \langle    a(t) \rangle    $ is the number of condensed magnons 
and $ \vartheta  (t)$ denotes the phase (Fig. \ref{fig:MagnonSpin}; we recall the linearized transformation \cite{HP} $ S^+  \approx  \sqrt{2S} a$).
The ODLRO becomes zero when magnons are not in condensation.
The quasiequilibrium magnon condensate is realized not in the ground state but in the metastable state, 
indicating that the phase becomes time-dependent,
\begin{eqnarray}
  \frac{d \vartheta  (t)}{dt}  \not=  0.
\end{eqnarray}
In such a quasiequilibrium condensate phase, all spins precess with the same finite frequency  $ d \vartheta (t)/dt $ and can be identified with a macroscopic coherent spin precession. See Figs.  \ref{fig:MagnonSpin} and  \ref{fig:Incoherent}.

Note that throughout this paper, we use the terminology `magnon coherent state' for the state which gives a nonzero value to the ODLRO $\langle a \rangle  \not=0$.
In terms of the spin variables, $\langle a \rangle \equiv  \sqrt{N_{\rm{cond.}}}  {\rm{e}}^{i \vartheta  } \not=0$ means that all spins precess with the same frequency (i.e., {\it{coherently}}) and a constant phase difference, which we call macroscopic {\it{coherent}} spin precession; such coherent spin precession might be identified with a precession of a macroscopic spins and therefore, it could be treated {\it{semiclassically}} (see also Sec. \ref{subsec:outlook}).
See Refs. [\onlinecite{Glauber,mehta}] for the time-evolution of coherent states.

%%%%%%%%%%%%%%%%%%%%%%%%%%%%%%%%%%%%%%%%%%%%%%%
\begin{table*}
\caption{
\label{tab:magnon}
Properties of noncondensed magnons and those of magnons in quasiequilibrium condensation.
They are distinguished by the ODLRO $ \langle a \rangle   $ associated with a macroscopic coherent state.
Phases play the key role on the transport of magnons in quasiequilibrium condensation and a magnetic field difference $\Delta B \not= 0 $ leads to ac Josephson currents, while dc currents when magnons are not in condensation. This is a good platform to experimentally observe condensed magnon currents. 
The quasiequilibrium magnon condensate is realized not in the ground state but in the metastable state, where all spins precess with the same frequency $d \vartheta  (t)/(dt) \not=  0$ and can be identified with a macroscopic coherent spin precession. 
The number of magnons in quasiequilibrium condensation and the time-evolution can be experimentally \cite{demokritov,ultrahot,SergaBEC} investigated by Brillouin light scattering spectroscopy \cite{demokritovReport}, and at room temperature, the number density of noncondensed magnons is \cite{demokritov} much larger than that of such condensed magnons.
}
%%%%%%%%%%%%
\begin{ruledtabular}
\begin{tabular}{cccccc}
%%%%%%%%%%%%%%%%%%%%%%%%%%%%%%%%%%%%%%%%%%%%\cite{demokritov,ultrahot,SergaBEC}
    Noncondensed magnon (i.e., thermally-induced magnon \cite{bunkov,Zapata}) &  Quasiequilibrium condensed magnon   \\ \hline
%%%%%%%%%%%%%%%%%%%%%%%%%%%%%%%%%%%%%%%%%%%%
$\langle a \rangle =0 $    &   $\langle a \rangle \not= 0 $  \\
 Individual magnon   & Macroscopic coherent magnon state \\
   Incoherent spin precession &   Macroscopic coherent spin precession   \\
   Fig. \ref{fig:Incoherent} (c) &    Fig. \ref{fig:Incoherent} (a) and (b)   \\
   $\langle a^{\dagger } a \rangle \sim  \int d \omega  ({\rm{e}}^{\beta \omega }-1)^{-1} $ 
& $N_{\rm{cond.}} =\langle    a(t) \rangle ^{\ast }  \langle    a(t) \rangle    $  
with $\langle a (t) \rangle = \sqrt{N_{\rm{cond.}}} {\rm{e}}^{i\vartheta (t)} $   \\
  Sum over various low-energy modes & A macroscopic number of magnons occupies a single state  \\
 Magnon current (i.e., spin-wave spin current \cite{spinwave})  &  Condensed magnon current \cite{MagnonSupercurrent}  \\
Number density \cite{demokritov} at room temperature; $ 10^{21}  $-$ 10^{23}  $cm$^{-3}$  
&  Number density \cite{demokritov} at room temperature; $ 10^{18}  $-$ 10^{19}  $cm$^{-3}$      \\
 A dc current $ {\cal{O}} (J_{\rm{ex}}^2)$ by $\Delta B \not= 0 $ in the junction (Fig. \ref{fig:WF2})
& An ac current $ {\cal{O}} (J_{\rm{ex}})$ by $\Delta B \not= 0 $ in the junction (Fig. \ref{fig:WF2})  \\
 Voltage drop $V_{\rm{m}}$ in Fig. \ref{fig:E}:   $V_{\rm{m}} \sim $mV  &  Voltage drop $V_{\rm{m}}$ in Fig. \ref{fig:E}:   $V_{\rm{m}} \sim $nV   \\
\end{tabular}
\end{ruledtabular}
\end{table*}
%%%%%%%%%%%%%%%%%%%%%%%%%%%%%%%%%%

%%%%%%%%%%%%%%%%%%%%%%%%%%%%%%%%%
\subsubsection{Magnon Josephson Equation}
\label{subsubsec:MJE}
%%%%%%%%%%%%%%%%%%%%%%%%%%%%%%%%%

The ODLRO is analogous to the order parameter for the conventional superconductors (Table \ref{tab:SCBEC}). 
Therefore in analogy to the superconductors, we \cite{KKPD} can discuss the condensed magnon transport in the magnetic insulating junction (Fig. \ref{fig:WF3}) and found the Josephson effects.

Assuming $T_{\rm{L}} =T_{\rm{R}}=0$, the quasiequilibrium magnon condensates in the junction are characterized by
\begin{eqnarray}
   \langle    a_{\rm{L(R)}}(t) \rangle  =  \sqrt{n_{\rm{L(R)}}(t)}  {\rm{e}}^{i \vartheta _{\rm L(R)} (t)},
\end{eqnarray}
where the variable $n_{\rm{L(R)}}(t)$ represents the number density of condensed magnons in the left (right) FI 
and $\vartheta_{\rm{L(R)}}(t)   $ denotes the phase (see Ref. [\onlinecite{KKPD}] for the Gross-Pitaevskii Hamiltonian derived from the original spin Hamiltonian).
The magnon population imbalance and the relative phase are defined by
\begin{subequations}
\begin{eqnarray}
z(t) &  \equiv & [n_{\rm{L}}(t) - n_{\rm{R}}(t)]/n_{\rm T},   \\
\theta(t) &  \equiv  & \vartheta _{\rm R} (t)  -  \vartheta _{\rm L} (t),
\end{eqnarray}
\end{subequations}
where the constant  $n_{\rm T} \equiv  n_{\rm L}(t) + n_{\rm R}(t)$ denotes the total population in the junction. 
After switching off the microwaves, the $U(1)$-symmetry of the system is recovered and the number of condensed magnons may be assumed to be conserved at zero temperature.
%%%%%%%%%%%%%%%%%%%%%%%%%%%%%%%%%%%%%%%%%%%%%%
An external electric field $   {\mathbf{E}} = E{\bf e}_y $ is applied to the interface.
Consequently, during the tunneling process, magnons acquire the A-C phase (Table \ref{tab:BerryPhase}) and it is described by
\begin{eqnarray}
{\cal{H}}_{\rm{ex}}^{\rm{A{\mathchar`-}C}}  &=& -J_{\rm{ex}} S  \sum_{\langle \Gamma_{\rm{L}} \Gamma_{\rm{R}} \rangle}
                                                 (a_{\Gamma_{\rm{L}}} a_{\Gamma_{\rm{R}}}^{\dagger }  {\rm{e}}^{- i  \theta _{\rm{A{\mathchar`-}C}}}
                                              + {\rm{h.c.}}),
                                              % + a_{\Gamma_{\rm{L}}}^{\dagger } a_{\Gamma_{\rm{R}}}    {\rm{e}}^{ i  \theta _{\rm{A{\mathchar`-}C}}}  ),
 \label{eqn:ac}
\end{eqnarray}
where $  \theta _{\rm{A{\mathchar`-}C}} = [g \mu_{\rm{B}}/(\hbar c^2)] E \Delta x$ for the geometry under consideration and now
$ \Delta x $ is the distance between boundary spins.
This A-C effect gives a handle to electromagnetically control magnon transport in the junction.
Note that such an effect on magnons has been experimentally observed recently in Ref. [\onlinecite{ACspinwave}].
%%%%%%%%%%%%%%%%%%%
In terms of the canonically conjugate variables $z(t)$ and $ \theta (t)$, the condensed magnon transport in the junction is described by
\begin{subequations}
\begin{eqnarray}
       \frac{ d z }{d {\cal{T}} }   &=&  -   \sqrt{1- z^2} {\rm{sin}} (\theta + \theta _{\rm{A{\mathchar`-}C}}),     \label{eqn:twostate5}    \\ 
       \frac{ d \theta  }{d {\cal{T}} }   &=&   \Delta  {\cal{E}} + \Lambda  z  
                                             +    \frac{ z}{\sqrt{1-z^2}}  {\rm{cos}} (\theta + \theta _{\rm{A{\mathchar`-}C}}),
   \label{eqn:twostate6}
\end{eqnarray}
\end{subequations}
where
\begin{subequations}
\begin{eqnarray}
       {\cal{T}}   &\equiv &  2 K_0 t/\hbar,     \\
       \Delta  {\cal{E}}  &=& \frac{{\cal{E}}_{\rm{L}}- {\cal{E}}_{\rm{R}}}{2 K_{0}} 
                                           + \frac{U_{\rm{L}}-U_{\rm{R}}}{4 K_0} n_{\rm{T}}, \label{eqn:parameter}     \\
       \Lambda  &=&  \frac{U_{\rm{L}} + U_{\rm{R}}}{4 K_0} n_{\rm{T}},
   \label{eqn:parameter2}
\end{eqnarray}
\end{subequations}
with
\begin{subequations}
\begin{eqnarray} 
  {\cal{E}}_{\rm{L(R)}} & = & 4 JS  (1   -   \eta  )  +g\mu_{\rm{B}}  B_{\rm{L(R)}},  \label{eqn:E} \\
 U_{\rm{L(R)}} &=&  -2J(1-\eta) a^3.   \label{eqn:U}         
 \label{eqn:K}
\end{eqnarray}
\end{subequations}
The tunneling amplitude is defined by $K_0  \equiv  J_{\rm ex} S$  
and magnon-magnon interactions arise \cite{KPD} from the spin anisotropy $ 0 < \eta \not=1$ [Eq. (\ref{eqn:U})] of
the spin Hamiltonian  $  {\cal{H}}_{\rm{H}}  $  for a single FI given by ($J <0$)
   $ {\cal{H}}_{\rm{H}}  =  \sum_{\langle i j\rangle} {\bf S}_i \cdot {\bf J} \cdot {\bf S}_j -  g\mu_{\rm{B}}  {\bf B} \cdot \sum_i {\bf S}_i$,
where ${\bf J}$ denotes a diagonal $3\times 3$-matrix with $\textrm{diag}({\bf J}) = J\{ 1, 1,\eta\}$. 
%%%%%%%%%%%%%%%%%%%%%%%%%%%%%%%%%%%%%%%%%%%%%%%%%%%%%%%
Eqs. (\ref{eqn:twostate5}) and (\ref{eqn:twostate6}) are the magnon Josephson equations:
Eq. (\ref{eqn:twostate5}) describes the magnon Josephson current and Eq. (\ref{eqn:twostate6}) the time-evolution of the relative phase,
which mean that the phases $\vartheta _{\rm R} (t)$ and $\vartheta _{\rm L} (t)$ play the key role on magnon transport in quasiequilibrium condensation.
The magnon Josephson current arises from terms of order $ {\cal{O}} (J_{\rm{ex}})$, 
which is in sharp contrast to the noncondensed magnon current $ {\cal{O}} (J_{\rm{ex}}^2)$ in the junction (see Tables \ref{tab:magnon} and \ref{tab:SCBEC}).
%%%%%%%%%%%%%%%%%%%%%%%%%%%%%%%%%%%%%%%%%%%%%%%%%%%%%%%
Thus the condensed magnon transport in the junction is described by the magnon Josephson equations.
We note for condensed magnons, noncondensed magnons work as an effective magnetic field 
and such an effect can be taken into account in $B_{\rm{L(R)}}$ [Eq. (\ref{eqn:E})].

\begin{figure}[h]
\begin{center}
\includegraphics[width=8.5cm,clip]{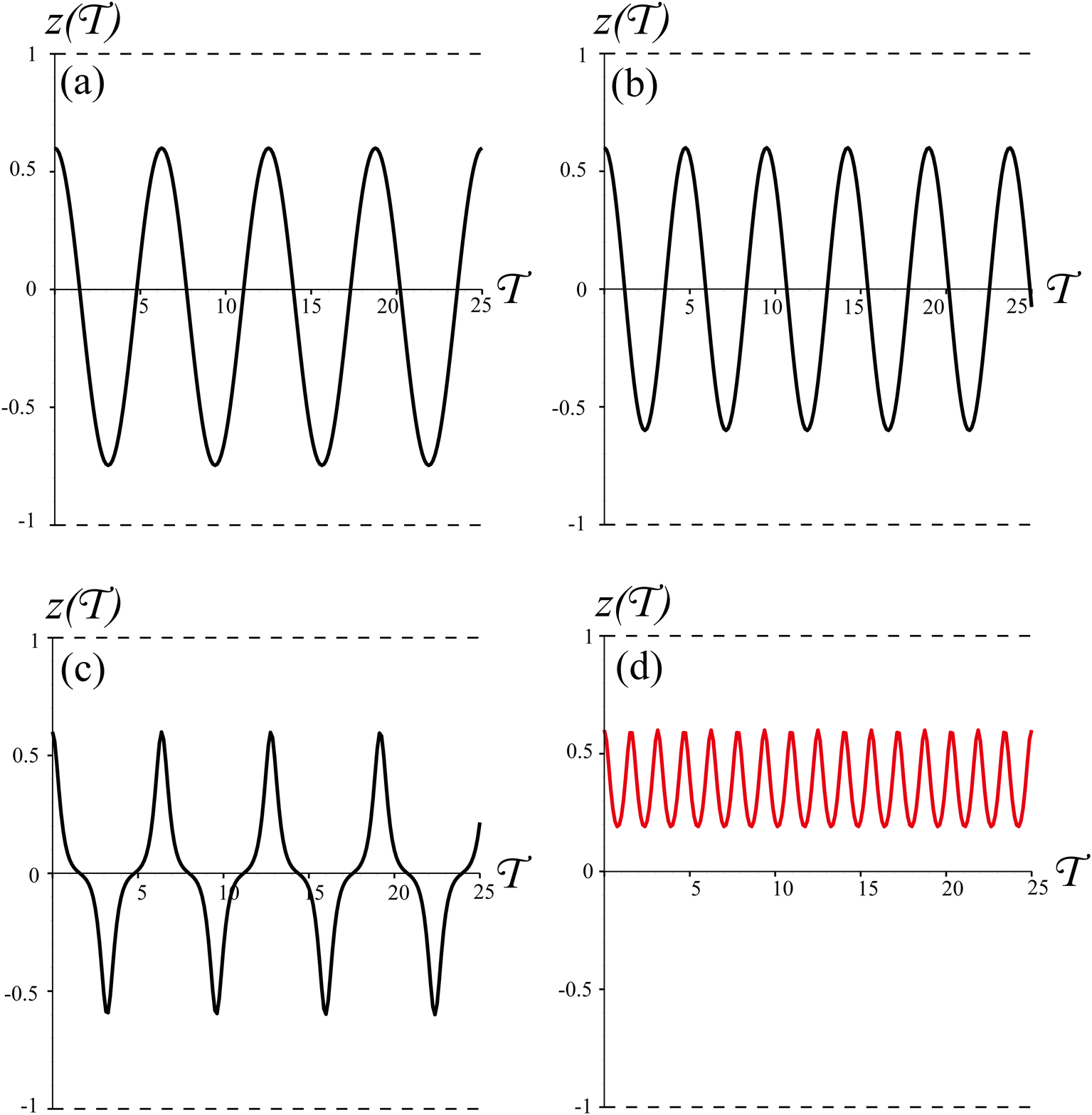}
\caption{(Color online)
The ac Josephson effects: Plots of the population imbalance $z({\cal{T}})$ as function of the rescaled time ${\cal{T}}= (2  J_{\rm{ex}} S /\hbar) t $.
The rescaled time $ {\cal{T}} =1 $ corresponds to $t = 1$ ns for $ J_{\rm{ex}} = 0.25$ $\mu$eV and $S=2$.
(a)  $ \Delta {\cal{E}} = 0.1 $,   $ \Lambda =0 $, $  \theta _{\rm{A{\mathchar`-}C}}   =0 $, $ z(0) =0.6 $, and $ \theta  (0) =0  $. 
The period of an oscillation is $ {\cal{T}} =  6$ ns.
(b) - (d)  $ \Delta {\cal{E}} =  \theta _{\rm{A{\mathchar`-}C}}  = 0 $, $ z(0) =0.6 $, and $ \theta  (0) =0  $,
which give $ \Lambda _{\rm{c}}   = 10$.
(b) $ \Lambda =1$,
(c) $ \Lambda =9.99$, and
(d) $ \Lambda =11 $: The MQST occurs since  $  \Lambda  > \Lambda_{\rm{c}}$.
Reprinted with permission from Ref. [\onlinecite{KKPD}].  
\label{fig:acdcMQST} }
\end{center}
\end{figure}

Fig. \ref{fig:acdcMQST} shows a numerical plot of ac Josephson effect and that of a macroscopic quantum self-trapping (MQST).
The self-interaction $\Lambda \not=0$ results from spin anisotropies $ \eta \not=  1$ and characterizes the period of ac Josephson effects.
A MQST occurs \cite{smerzi,smerzi2,smerzi3,albiez} when the value satisfies 
$  \Lambda  > \Lambda_{\rm{c}}$, 
where 
\begin{eqnarray}
         \Lambda _{\rm{c}}
= \frac{ 1+  \sqrt{1-z(0)^2}    {\rm{cos}} ( \theta(0) + \theta _{\rm{A{\mathchar`-}C}})   }{ z(0)^2/2   }.
   \label{eqn:critical}
\end{eqnarray}
On the other hand, in the isotropic case $ \eta =  1$ (i.e., $\Lambda  = 0 $),
ac Josephson effects become characterized by the nonlinear effect  
$ ({ z}/{\sqrt{1-z^2}})  {\rm{cos}} (\theta + \theta _{\rm{A{\mathchar`-}C}})$ in Eq. (\ref{eqn:twostate6}).
%%%%%%%%%%%%%%%%%%%%
We numerically found that the period is determined by the nonlinear effect and little influenced by the magnetic field difference $\Delta {\cal{E}} $
since the maximum value of magnetic field gradient within experimental reach is estimated by $\partial B/(\partial x) \leq   1$T/cm,
which results in $ \Delta {\cal{E}}  \leq   1 $.
Even without any magnetic field differences, ac Josephson effects are generated by the nonlinear effect with an initial population imbalance $ z(0) \not=0$.
The period $t_{\rm{ac}} $ is estimated by $t_{\rm{ac}} = 3\hbar /(J_{\rm{ex}}S)$, 
which becomes within experimental reach \cite{DemokritovPrivate} $t_{\rm{ac}}  =10 $ns when $ J_{\rm{ex}} = 0.03 \mu $eV and $S=10$ as an example.
Such a weak tunneling amplitude may be realized by inserting a thin nonmagnetic insulator (NI) between the FIs 
and realizing the multi-layered Josephson junction FI/NI/FI.

\begin{figure}[h]
\begin{center}
\includegraphics[width=8.5cm,clip]{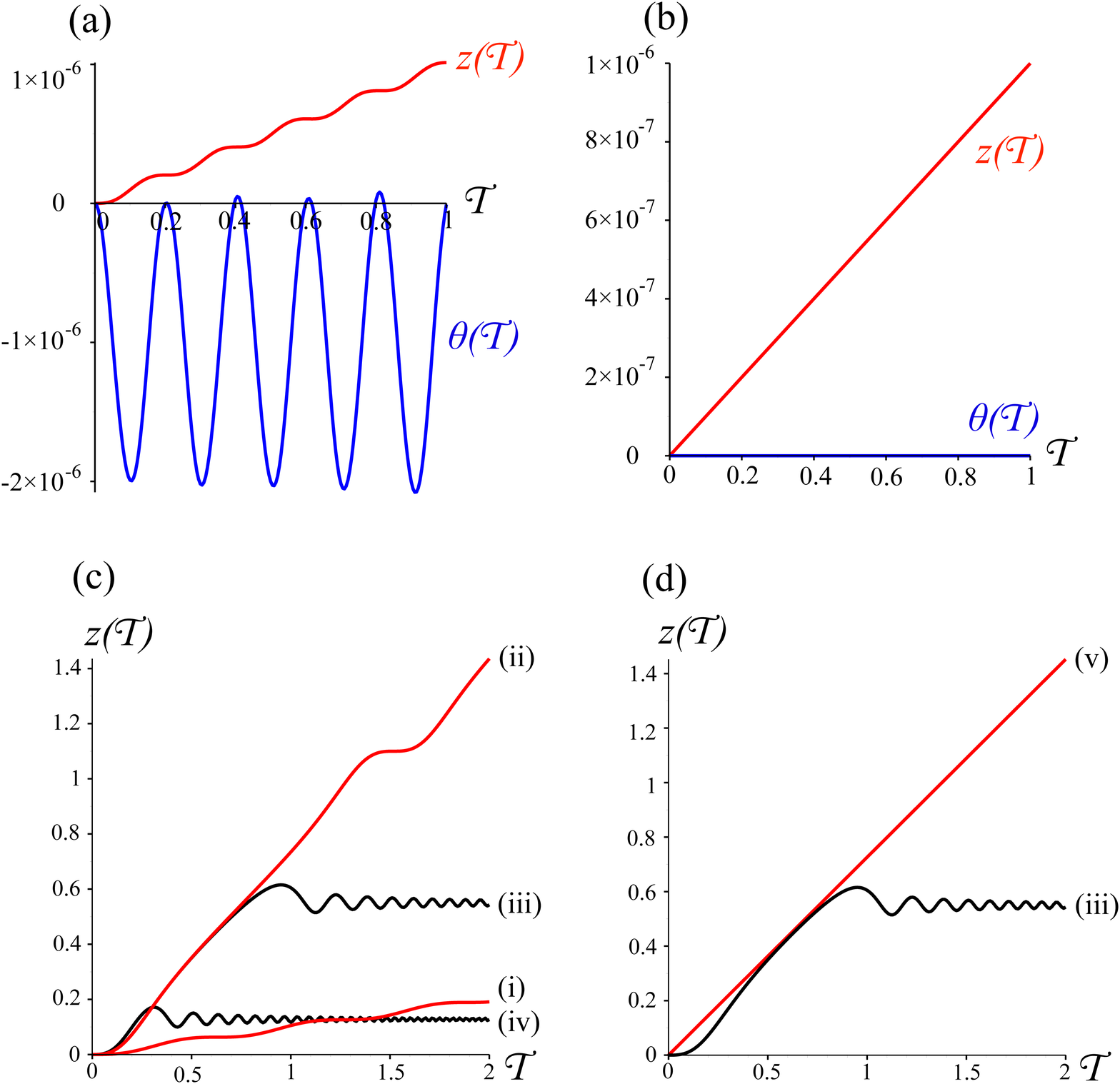}
\caption{(Color online)
The dc Josephson effects: Plots of the population imbalance $z({\cal{T}})$ and the relative phase $\theta({\cal{T}})$ 
as function of the rescaled time ${\cal{T}}$.
for (a) (b)  $ {z} _0 =  10^{-6}  $, $ \Lambda =  10^{3} (\gg  1)$, and $z(0)=\theta (0) = 0$. 
(a) $\theta _{\rm{A{\mathchar`-}C}}  = 0$.
(b) $ \theta _{\rm{A{\mathchar`-}C}}  =  -\arcsin({z}_0)=-10^{-6}$.
(c) The breakdown of the dc Josephson effect ($\theta_{\rm{A{\mathchar`-}C}} =0$) due to increase of $b_0$. 
The transition between the dc region [(i), (ii)] and the ac one [(iii), (iv)] takes place for  $ {z} _0 \approx    0.725 $ due to the absence of the A-C phase. 
On the conditions $(z(0), \theta (0)) =(0, 0)$ and $  \Lambda =100 $  ($ \gg  1 $), each $ {z} _0 = {b}_0  / \Lambda $  reads 
(i) 0.100, (ii) 0.724, (iii) 0.726, and (iv) 1.100. 
(d) Example of the recovery of the dc Josephson effect from the ac effect through the A-C phase in the region $ {z} _0 \leq 1 $   (i.e., $ {b}_0  \leq  \Lambda $).
(iii) $\theta_{\rm{A{\mathchar`-}C}} =0$, (v) $\theta_{\rm{A{\mathchar`-}C}} = -\arcsin({z}_0)$.
Reprinted with permission from Ref. [\onlinecite{KKPD}].  
\label{fig:acdcTransition} }
\end{center}
\end{figure}

Fig. \ref{fig:acdcTransition} shows numerical plots of a dc Josephson effect and a transition between the ac and dc Josephson effects.
A dc Josephson effect is generated by a time-dependent magnetic field
whose magnitude increases over time with a rate $b_0$ for a limited rescaled time ${\cal{T}}_0$;
\begin{equation}
     \Delta {\cal{E}}({\cal{T}}) = \frac{g \mu_{\rm{B}} }{2 K_0} \left(B_{\rm{L}}  -B_{\rm{R}}\right) 
= \left\{ \begin{array}{cc} - b_0 {\cal{T}} & \textrm{ for }  {\cal{T}} \in (0, {\cal{T}}_0), \\ 
0 & \textrm{otherwise}. \end{array} \right.
   \label{eqn:70}
\end{equation}
Assuming  $\Lambda \gg  1 $ and $|z|\ll 1$, the Josephson equations [Eqs. (\ref{eqn:twostate5}) and (\ref{eqn:twostate6})] reduce to
\begin{subequations}
\begin{eqnarray}
       \frac{ d z }{d {\cal{T}} }   &\approx &  - {\rm{sin}} (\theta  +\theta_{\rm{A{\mathchar`-}C}}),    \label{eqn:initial3}  \\
       \frac{ d \theta  }{d {\cal{T}} }   &\approx &   - {b}_0 {\cal{T}} + \Lambda z.
   \label{eqn:initial4}
\end{eqnarray}
\end{subequations}
The steady-state solution of dc effects  for a finite time  is given by 
\begin{equation}
z ({\cal{T}}) = {z} _0 {\cal{T}} \textrm{ and } \theta({\cal{T}}) = -\arcsin({z}_0) - \theta_{\rm{A{\mathchar`-}C}},  \label{eqn:KoukiKevin}
\end{equation}
with ${z} _0 \equiv   {b}_0  / \Lambda$.
%%%%%%%%%%%%%%%%%%%%%%%%%%%%%
Supposing $\theta(0) = 0$, we remark that unless $\theta_{\rm{A{\mathchar`-}C}}$ is tuned to the value
\begin{equation}
\theta_{\rm{A{\mathchar`-}C}} = -\arcsin({z}_0),  \label{eqn:A-CforDC}
\end{equation}
a mismatch in $\theta({\cal{T}})$ with the steady-state solution arises.
This leads to a phenomenon that the value of $ {z}_0$ for a transition between the ac and dc Josephson effects is reduced by a numerical factor $\approx 0.725$.
See Ref. [\onlinecite{KKPD}] for details.
%%%%%%%%%%%%%%%%%%%%%%%%%%%%%%%%%%

So far we have assumed bulk FIs where magnetic dipole-dipole interactions are negligible \cite{tupitsyn}. 
Such dipolar effects, however, become important in thin films, resulting in a modified dispersion for magnons 
where the lowest energy mode becomes $k =  k_{\rm{m}}  \sim  10^4 $/cm for YIG thin films \cite{demokritov,ultrahot,SergaBEC,LTP};
the magnitude of $k_{\rm{m}}$ depends on the width of YIG thin films.
Still, we confirmed \cite{noteJosephson} that the magnon Josephson effects qualitatively remain valid; 
the contribution is added into $ {\cal{E}}_{\rm{L(R)}}  $ in Eq. (\ref{eqn:E}) and it is modified by
$  {\cal{E}}_{\rm{L(R)}}  =  4 JS  (1   -   \eta  )  +g\mu_{\rm{B}}  B_{\rm{L(R)}} -J S (1 +   \eta  ) (a  k_{\rm{m}}^{\rm{L(R)}}) ^2$,
where $k_{\rm{m}}^{\rm{L(R)}}$ is the lowest energy mode in the left (right) FI.

Lastly, we remark that when $  \theta _{\rm{A{\mathchar`-}C}} =   0$ in Eqs. (\ref{eqn:twostate5}) and (\ref{eqn:twostate6}), 
the description mathematically reduces to the one for cold atoms \cite{smerzi,smerzi2,smerzi3,Zapata,LeggettReview},
where the Josephson effects and the MQST have already been observed \cite{albiez,Levy} experimentally.
Similar equations for $  \theta _{\rm{A{\mathchar`-}C}} =   0$ have been proposed phenomenologically 
for ferromagnets \cite{troncoso2} and for antiferromagnets \cite{Schilling}.
%%%%%%%%%%%%%%%%%%%%%%%%%
It should be noted \cite{bunkovprivate} that {\it{Bose-Einstein condensation}} of magnons and the resulting several phenomena ({\it{e.g.}}, Josephson effects) in Helium-3 have already been intensively investigated by Bunkov-Volovik \cite{bunkov}, and the microscopic mechanisms are well understood both theoretically and experimentally. See Refs. [\onlinecite{bunkovQ,bunkovJETP}] as an example and the review article [\onlinecite{bunkov}] for details of their works on Helium-3.

%%%%%%%%%%%%%%%%%%%%%%%%%%%%%%%%%%%%%%%%%%%%%%%
\begin{table*}
\caption{
\label{tab:SCBEC}
Magnon analogues of the conventional superconductors characterized \cite{leggett} by the order parameter for copper pairs,
$\langle  c_{{\mathbf{k}}\uparrow }  c_{-{\mathbf{k}}\downarrow  } \rangle$, where $c$ is an annihilation operator for fermions.
Josephson effects in junction of two bulk superconductors are characterized by the canonically conjugate variables; the number imbalance $  \Delta {\cal{N}}   $ and the relative phase $ \Delta  \varphi $.
The tunneling coupling $J_{\rm{SC}}$ arises from a finite overlap of the wave functions and characterizes the critical current 
$ I_{\rm{c}}=  e^{\ast }J_{\rm{SC}}/\hbar   $.
The time-evolution of the relative phase is produced by an external voltage ${\cal{V}}$ across the superconducting junction,
while the role is played by the magnetic field difference $\Delta  {\cal{E}}$, the magnon-magnon interactions $\Lambda $, and the nonlinear effect in magnon Josephson junctions. Therefore even without any magnetic field differences, ac magnon Josephson effects can be generated by the nonlinear effect under an initial population imbalance $ z(0) \not=0$.  See Ref. [\onlinecite{KKPD}] for the Gross-Pitaevskii \cite{hickbec} Hamiltonian of magnon condensates derived directly from the original spin Hamiltonian.
}
%%%%%%%%%%%%
\begin{ruledtabular}
\begin{tabular}{cccccc}
%%%%%%%%%%%%%%%%%%%%%%%%%%%%%%%%%%%%%%%%%%%
  & Conventional superconductor \cite{leggett} & Quasiequilibrium magnon condensate  \\ \hline
%%%%%%%%%%%%%%%%%%%%%%%%%%%%%%%%%%%%%%%%%%%
Order parameter 
&  $ \langle  c_{{\mathbf{k}}\uparrow }  c_{-{\mathbf{k}}\downarrow  } \rangle  \not= 0  $   
&   $ \langle  a(t) \rangle  =  \sqrt{N_{\rm{cond.}}}  {\rm{e}}^{i \vartheta (t)} \not= 0 $ with $d \vartheta  /(dt)  \not=  0$  \\
Carrier   &    $ e^{\ast }  \equiv   2e $   &     $ g\mu_{\rm{B}}$  \\
ac and dc Josephson effects & Possible \cite{Josephson}:  &  \ Possible \cite{KKPD} (see Sec. \ref{subsubsec:MJE} for MQST):  \\
 \   \  \ Current Josephson Eq. 
&  $ \frac{ d  }{d t }  \Delta {\cal{N}}(t)  = (2J_{\rm{SC}}/\hbar) {\rm{sin}}\Delta \varphi  (t)   $
&  $ \frac{ d z }{d {\cal{T}} }   =  -   \sqrt{1- z^2} {\rm{sin}} (\theta + \theta _{\rm{A{\mathchar`-}C}})$  \\
Phase Josephson Eq.  
& $ \frac{ d  }{d t }  \Delta \varphi (t)  = - 2 e {\cal{V}}(t)/\hbar    $
&  $\frac{ d \theta  }{d {\cal{T}} }   =   \Delta  {\cal{E}} + \Lambda  z  +  \frac{ z}{\sqrt{1-z^2}}  {\rm{cos}} (\theta + \theta _{\rm{A{\mathchar`-}C}})$ \\
Persistent current in Ring &  \  Possible:  & `Possible' \cite{KKPD,KPD} as long as in condensation:  \\
&   For \cite{kittel} over $10^{10^7} $s &   For \cite{demokritov,ultrahot,SergaBEC} about a few hundreds ns \\
Quantization in Ring:  & $ \Phi  = p \Phi _0 $: &    $\phi = p \phi _0$: \\
 (See Table \ref{tab:BerryPhase})  & $\Phi  \equiv   \oint_{\rm{Ring}}   \textrm{d} {\bf l} \cdot  {\bf A}$   &  
$ \phi  \equiv  \oint_{\rm{Ring}}   \textrm{d} {\bf l} \cdot   ({\bf E} \times  {\bf e}_z) $    \\
  &  $ \Phi _0 \equiv h c /e^{\ast } $  &  $\phi_0 \equiv  h c^2/(g\mu_{\rm{B}})$   \\
\end{tabular}
\end{ruledtabular}
\end{table*}
%%%%%%%%%%%%%%%%%%%%%%%%%%%%%%%%%%

\begin{figure}[h]
\begin{center}
\includegraphics[width=8cm,clip]{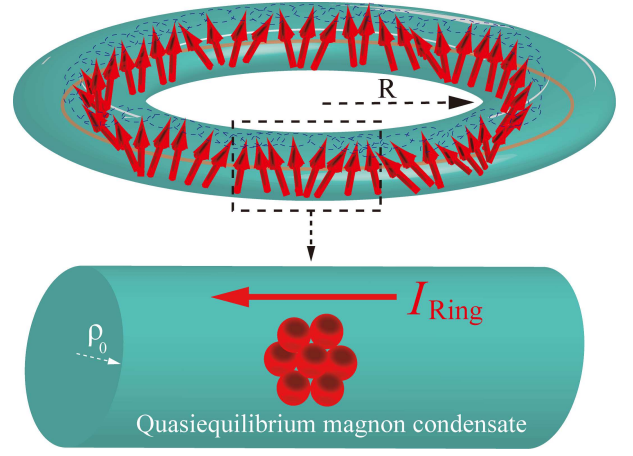}
\caption{(Color online)
Schematic representation of a condensed magnon ring with the radius $R$ and the enlarged view of the cylindrical wire that forms the ring. The radius of the wire is $ \rho_0 $.
A-C effects generate phase differences between the individual spins, but still they precess with the same frequency (i.e., coherently)  [Fig. \ref{fig:Incoherent} (b)].
Consequently, a condensed magnon current $   I_{\rm{Ring}}  $ flows `persistently' in the ring as long as magnons are in quasiequilibrium condensation.
Any persistent currents in the ground state are forbidden by Bloch theorem \cite{Yamamoto} and such a persistent current is possible since quasiequilibrium magnon condensates are not the ground state but a metastable state.
%%%%%%%%%%%%%%%%%%%%%%%%%%%%%%
The current is not steady when the quantization condition [Eq. (\ref{eq:ringQuantum})] is not satisfied.
However, the non-steady variations of the current away from its nonequilibrium steady state are negligibly small. 
%%%%%%%%%%%%%
This setup was proposed in Ref. [\onlinecite{KKPD}] and reprinted with permission.
\label{fig:Ring} }
\end{center}
\end{figure}

%%%%%%%%%%%%%%%%%%%%%%%%%%%%%%%
\subsection{Persistent Current and Quantization in Ring}
\label{subsec:persistent}
%%%%%%%%%%%%%%%%%%%%%%%%%%%%%%%

In analogy to a superconducting ring \cite{SCring,ImryPersistent} (see Table \ref{tab:SCBEC}), 
we introduce a condensed magnon ring as sketched in Fig. \ref{fig:Ring} and discuss the `persistent' condensed magnon current by the A-C effect.
Similar setup was proposed in Refs. [\onlinecite{LossPersistent,LossPersistent2,Kopietz,wei2}].

Due to the single-valuedness of the wave function of condensed magnons (i.e., macroscopically precessing localized spins) around the ring, 
the electric flux ${\phi}  $ is quantized as
\begin{eqnarray}
{\phi}  = p \phi_0   \   \      {\rm{with} }   \   \     p \in  {\mathbb{Z}},
\label{eq:ringQuantum}
\end{eqnarray}
where the integer $p$ is the phase winding number of the closed path around the ring with
\begin{subequations}
\begin{eqnarray}
\theta _{\rm{A{\mathchar`-}C}}
&= &
 \frac{g \mu_{\rm{B}} }{\hbar c^2} \oint_{\rm{Ring}} \textrm{d} {\bf l} \cdot \left( {\bf E} \times {\bf e}_z\right)   \\
 %\label{eq:AC_SQUID2}  
&\equiv  & 2\pi \frac{\phi}{\phi_0},
\label{eq:ring}
\end{eqnarray}
\end{subequations}
and the electric flux quantum \cite{wei2} 
\begin{eqnarray}
\phi_0 \equiv  h c^2/(g\mu_{\rm{B}}).
\end{eqnarray}
Assuming that a ring of radius $R$ consists of the cylindrical wire whose cross-section is $ \pi  \rho_0 ^2$ with the radius $\rho_0 $ as sketched in Fig. \ref{fig:Ring},
the magnitude of the `persistent' condensed magnon current in the ring becomes  (see Ref. [\onlinecite{KKPD}] for details)
\begin{eqnarray}
\mid   I_{\rm{Ring}}  \mid  =   
2 {\pi}  g \mu _{\rm{B}}   \mid J/\hbar \mid   S  \rho_0^2 a  n_{\rm{cond.}} \mid  {\rm{sin}}[2 a   \phi /( R  \phi_0)] \mid , \nonumber  \\
\end{eqnarray}
where $n_{\rm{cond.}}$ is the number density of condensed magnons.
The condensed magnon current flows `{\it{persistently}}' as long as magnons are in condensation.
Such a persistent current is possible since quasiequilibrium magnon condensates are not the ground state but a quasiequilibrium metastable state.
Remember that any persistent currents in the ground state are forbidden by Bloch theorem \cite{Yamamoto} (see also Ref. [\onlinecite{HWoshikawa}] for it).
%%%%%%%%%%%%%%%%%%%%%%%%%%%%%%%%%%%%%%%%%%%%%%%%%%%%
We note that the current in the ring is not steady when the quantization condition [Eq. (\ref{eq:ringQuantum})] is not satisfied, but
these non-steady variations of the current away from its nonequilibrium steady state are small, on the relative order of $1/p \ll 1$ (typically, $ p \geq  {\cal{O}}(10)$).
This is in contrast to a superconducting ring where the magnetic field of the supercurrent itself compensates the variations.
%%%%%%%%%%%%%%%%%%%%%%%%

We \cite{KPD} point out the possibility that using microwave pumping, the condensed magnon current might flow `persistently' even at finite temperatures. 
Dissipations arise at finite temperature. 
Such detrimental effects, however, can be compensated \cite{bunkov} by magnon injection through microwave pumping 
where the pumping rate is larger than the dissipative decay rate. 
%%%%%%%%%%%%%%%%%%%%%%%%%%%%%%%%%

%%%%%%%%%%%%%%%%%%%%%%%%%%%%%%%%%%%%%%%%%%%%%%%
\begin{table*}
\caption{
\label{tab:electromagnetism}
Electromagnetism \cite{dipole,magnon2,KKPD} of the magnon current $I_{\rm{Ring}}$  in the cylindrical wire (see Figs. \ref{fig:Ring} and \ref{fig:E}).
A steady electron current produces a static magnetic field $B(r) $, while a steady spin current ({\it{e.g.}}, the persistent condensed magnon current $I_{\rm{Ring}}$ in the ring) generates a static electric field $E_{\rm{m}}(r) $,
which results in the voltage drop $V_{\rm{m}}$ between the points (i) and (ii) in Fig. \ref{fig:E}.
}
%%%%%%%%%%%%
\begin{ruledtabular}
\begin{tabular}{cccccc}
%%%%%%%%%%%%%%%%%%%%%%%%%%%%%%%%%%%%%%%%%%%%
   &  Electron current $I_{\rm{e}}$ &  Magnon current $I_{\rm{Ring}}$    \\   \hline
   Carrier    &     $e $       &      $ g \mu_{\rm{B}}  $    \\
 Driving forces  & Electric field and temperature gradient  & Magnetic field gradient \cite{Haldane2} and temperature gradient \\
Ampere$^{\prime}$s law   &  $B(r)  =  I_{\rm{e}}/(2 \pi r)$   &  $E_{\rm{m}}(r)  = \mu _0 I_{\rm{Ring}}/(2 \pi r^2 )$   \\
Ohm$^{\prime}$s law & $V = R   I_{\rm{e}}  $ &  $V_{\rm{m}} \approx     \mu _0  I_{\rm{Ring}}/(2 \pi r) $    \\
\end{tabular}
\end{ruledtabular}
\end{table*}
%%%%%%%%%%%%%%%%%%%%%%%%%%%%%%%%%%

\begin{figure}[h]
\begin{center}
\includegraphics[width=6cm,clip]{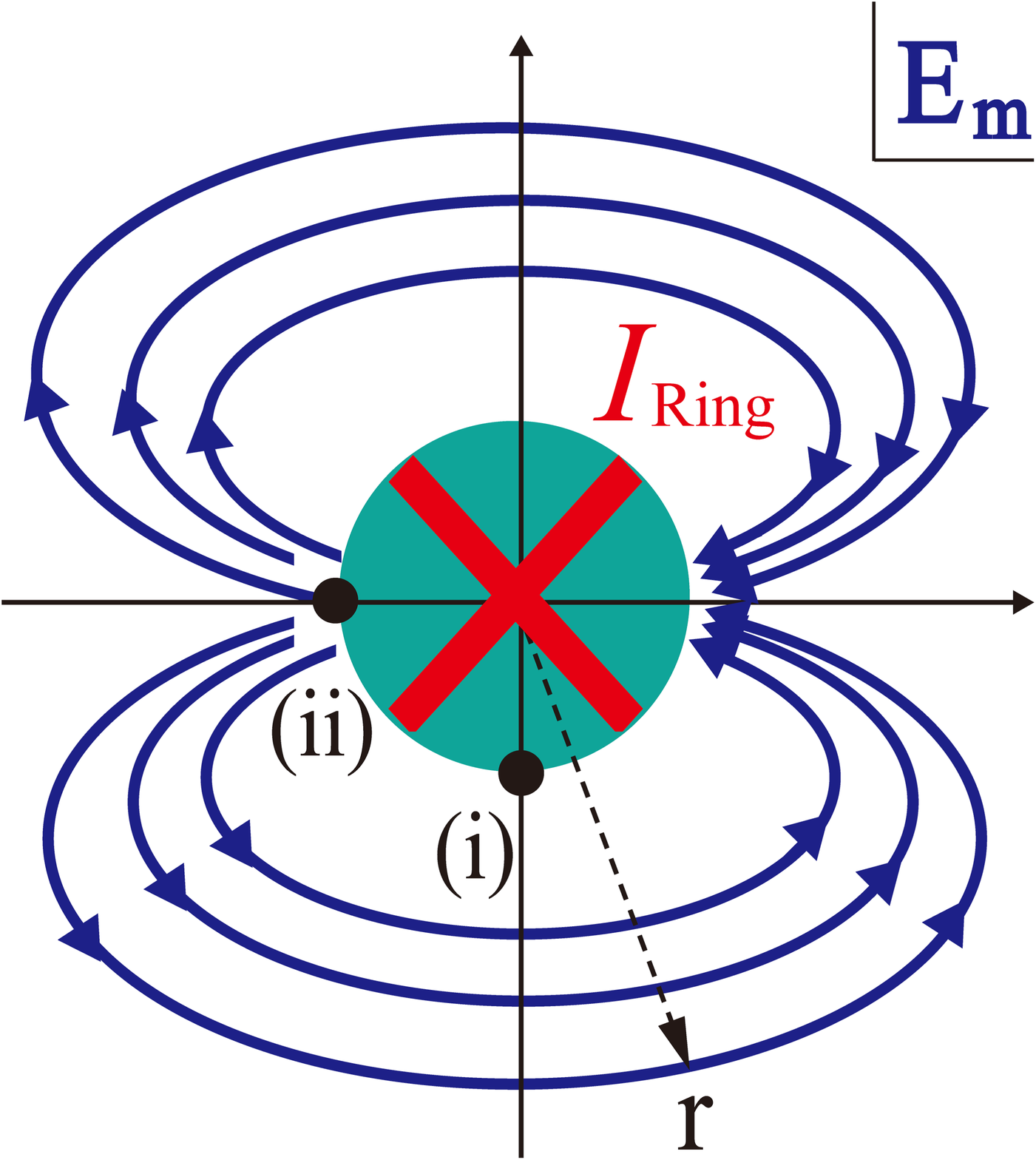}
\caption{(Color online)
A schematic picture of the cross-section of the cylindrical wire in Fig. \ref{fig:Ring}.
The persistent condensed magnon current  $   I_{\rm{Ring}} $ in the ring produces the electric field $  {\bf{E}}_{\rm{m}} (r)  $,
which leads to a measurable voltage difference  $ V_{\rm{m}}   $ between the points (i) and (ii).
Reprinted with permission from Ref. [\onlinecite{KKPD}].
\label{fig:E} }
\end{center}
\end{figure}

%%%%%%%%%%%%%%%%%%%%%%%%%%%
\subsection{Electromagnetism of Magnon Current}
\label{subsec:Measurement}
%%%%%%%%%%%%%%%%%%%%%%%%%%%

Magnons (i.e., spins) are magnetic dipoles, which can be regarded \cite{dipole} as a pair of oppositely charged magnetic monopoles of charge $\pm  q_{\rm{m}}$ separated by a distance $d$ in the limit $q_{\rm{m}}\rightarrow  \infty $, $d \rightarrow  0$ with $q_{\rm{m}} d = g\mu _{\rm{B}} $ fixed.
Taking this into account, Maxwell$^{\prime}$s equation can be formally enlarged.
Thus, based on the resulting correspondence between electricity and  magnetism, we theoretically proposed how to directly measure magnon currents without converting them into charge currents by the inverse spin Hall effect \cite{ISHE1,spinwave,phonon,uchidametal,uchidainsulator,mod2,adachireport}.

As is well-known, a steady charge current produces a static magnetic field.
By contrast, the electromagnetic consequence of a steady spin current is a static electric field (Table \ref{tab:electromagnetism}).
Spin currents produce two electric fields, one from the monopole and the other from the antimonopole.
Each magnetic monopole current produces a static, asymptotically dipolar, electric field.
Combining these two fields and taking the limit $d \rightarrow  0$, 
the resulting electric fields ${\bf{E}}_{\rm{m}} $ from magnon currents are shown in Fig. \ref{fig:E} for the cylindrical wire (Fig. \ref{fig:Ring}) as an example.
The magnitude  \cite{magnon2} can be estimated by
\begin{eqnarray}
 |{\bf{E}}_{\rm{m}} (r)  |= \frac{\mu _0}{2 \pi  r^2} \mid   I_{\rm{Ring}}   \mid,
\end{eqnarray}
which results in the voltage difference $V_{\rm{m}}$ between the points (i) and (ii).
Within experimentally realizable sample values, it can amount to the nV range for condensed magnon currents, while to the mV range for noncondensed magnon currents (see Refs. [\onlinecite{KKPD}] and [\onlinecite{magnonWF}] for details).
Although small, such values are within experimental reach.
%%%%%%%%%%%%%%%%%%%%%%%%%%%%
We remark that the resulting voltage drop from noncondensed magnon is about $ 10^6$ times larger than the one from condensed magnons 
since at room temperatures, the number density of noncondensed magnons is indeed \cite{demokritov}  much larger than that of condensed magnons 
(Table \ref{tab:magnon}).

%%%%%%%%%%%%%%%%%%%%%%%%%%%
\section{Concluding Remarks}
\label{sec:remark}
%%%%%%%%%%%%%%%%%%%%%%%%%%%

Using the spin-wave picture, we have classified magnon states in terms of their off-diagonal long-range order and  reviewed their resulting transport properties in insulating magnets. 
Despite their different statistics, we found many  phenomena analogous to electron transport in metals: a Wiedemann-Franz law for magnon transport, some Onsager reciprocal relation between the magnon Seebeck and Peltier coefficients,  Hall effects of magnons, a quasiequilibrium magnon condensation, the ac and dc magnon Josephson effects, a quantized persistent current in Aharonov-Casher effects, a magnetic analog of a quantum RC circuit,  magnon transistors, {\it{etc}}.
%%%%%%%%%%%%%%
Experimentally observing these phenomena would undoubtedly constitute milestones in this research area.
We believe these goals are within reach using the present experimental techniques \cite{demokritov,ultrahot,SergaBEC,MagnonSupercurrent,demokritovReport,MagnonPhonon,ACspinwave,MagnonC}.

%Finally, we apologize to the many whose work we were not able to treat due to the page limit on the number of this review article.

%%%%%%%%%%%%%
\section{Outlook}
\label{sec:outlook}
%%%%%%%%%%%%%
%[Dear Daniel] With open issues, I have remarked several issues on spin-superfluid, spin-supercurrent, quantum nature of BEC. Please check them. Please check the treatment of the report on the experimental realization of magnon-supercurrent from Hillbrand group.

Toward the next step of  magnonics \cite{MagnonSpintronics,ReviewPRapply,magnonics,spincal,spincalreview,caloritronics}, we enumerate fundamental open issues and provide some perspectives.

%%%%%%%%%%%%%
\subsection{Genuine Quantum Nature of Magnon}
\label{subsec:outlook}
%%%%%%%%%%%%%\cite{GluonBEC}
%\ps{Is the latter statement correct. I thought Kopietz reinterpreted the Magnon BEC of Serga et al. as a classical coherent object which is slightly different that saying that a BEC is a semi-classical object ?}
%Kouki: Respecting their work, I have changed into ``classical'' and revised accordingly. I had thought semi-classical includes classical; quantum-mechanics includes classical-mechanics.

According to their bosonic nature magnons can form condensate and using microwave pumping, it can be realized also as a metastable state \cite{demokritov,ultrahot,SergaBEC,bunkov,Yukalov,BatistaBEC,Mills} even at room temperature; applying microwave, the system is driven into a nonequilibrium steady state and the zero-mode of magnons continues to be injected. After switching off the microwave, toward the true lowest energy state, the system undergoes a thermalization (i.e., relaxation) process subject to magnon-magnon interactions and thereby reaches a metastable state where the pumped magnons form condensate. 
Such dynamic or kinetic condensation  in quasiequilibrium might be viewed as a {\it{classical}} phenomenon \cite{KopietzBEC} due to the coupling with a thermal bath (i.e.,  thermally activated lattice vibration) \cite{Abragam,C-L,CL_text}; applying a microwave, spins precess coherently and behave like a macroscopic spin. Since such a macroscopic object is subject to thermal bath, the relaxation and {\it{decoherence}} time \cite{CL_text,Zurek,DL_QuantSpinMagn} of a quantum superposition of macroscopically distinct states becomes extremely short at room temperature
%much shorter than the relaxation time 
and it loses the quantum-mechanical properties in an extremely short time. 
The dynamics thus might be well described \cite{KopietzBEC} by Landau-Lifshitz-Gilbert equation\cite{Gilbert}.
%%%%%%%%%%%%%%%%%%%%
One may therefore wonder how to test and probe genuine quantum-mechanical properties \cite{DL_rev_QuantSpinDeco,DL_QuantSpinMagn} such as the {\it{quantum coherence}} \cite{Zurek} of such a macroscopic object. 
As to quantum coherence \cite{noteCoherence} effects in mesoscopic spin systems, see Refs. [\onlinecite{DL_rev_QuantSpinDeco,DL_QuantSpinMagn,DL_QuantCoherence}] and Ref. [\onlinecite{LeggettReview}] for those in the dilute atomic alkali gases.

As a  theoretical perspective \cite{Abragam,C-L,CL_text} we note that, at finite temperatures, lattice vibrations become thermally activated and generally couple to spin degrees of freedom in solids. Partially tracing out the phonon degrees of freedom from the spin-lattice interaction, the spin dynamics becomes being described by a reduced action and generally exhibits a non-unitary time-evolution (i.e. dissipation).
That is, as long as phonons are alive, dissipations and the decoherence are inevitable in solids.
How to overcome such detrimental effects and go beyond magnon injection by microwave pumping \cite{KPD}?
%%%%%%%%%%%%%%%%%%%%%%%
The conventional superconductors solve this issue by absorbing phonon degrees of freedom to form Cooper pairs. Could a similar mechanism be possible here?
Isolated quantum system \cite{TasakiHara,Kitagawa,Moessner,AokiTsujiOka} (i.e., low temperature) might offer a platform to explore the possibility of genuine quantum-mechanical condensation \cite{GluonBECnote}.

%%%%%%%%%%%%%%%%%
\subsection{Dirac Magnons}
\label{subsec:outlook3}
%%%%%%%%%%%%%%%%%

Magnons with a quadratic dispersion relation are identified with nonrelativistic-like magnons,
while those with a linear dispersion can be viewed as relativistic-like magnons. 
On cubic lattices, the magnetization of FMs is characterized by nonrelativistic-like magnons contrary to AFs which have  relativistic-like magnon excitations.
Fransson {\it{et al.}} \cite{DiracMagnon} have recently found that such a relativistic-like \footnote{Relativistic-like magnons on honeycomb lattices were discussed, for instance, also in Refs. [\onlinecite{zigzag}] and [\onlinecite{zigzag2}].} magnon can emerge from the geometric properties of honeycomb lattices (i.e., bipartite lattices) both in FMs and AFs.
In analogy with Dirac fermions in graphene \cite{Graphene}, the relativistic-like magnons are described by a magnon Dirac equation and may be called {\it{Dirac magnons}} (see Ref. [\onlinecite{DiracMagnon}] for details).
Though their quantum-statistical properties are different, some analogous phenomena to Dirac fermions are still expected to emerge in Dirac magnon systems.
Finding  intrinsic relativistic effects associated with Dirac magnons transport is desired.

To this end, one of the promising platforms is quantum Hall systems\cite{FQHEnote}.
The Hall conductivity of nonrelativistic-like fermions is \cite{TKNN,Kohmoto}
$\sigma _{xy}= - (e^2/h) \nu  $ with Chern number $\nu  \in  {\mathbb{Z}} $, while that of Dirac fermions becomes \cite{QHErelEl}
$\sigma _{xy}^{\rm{Dirac}}= - (2 e^2/h) (2\nu +1) $ due to the quantum anomaly \cite{QHErel1984,QHErel1985,QHErel1991} of the lowest Landau level ({\it{e.g.}}, which makes the integer quantum Hall effect in graphene  unconventional \cite{QHEgrapheneExp,GrapheneQHE}). Both Hall conductivity become discrete.
%%%%%%%%%%%%%%%%%%%%%%%%%%%%%%%%%%%%%%%%%%
Thus, quantum Hall effects are characterized by a topological invariant, a Chern number, associated with the Berry curvature.
Since the underlying Berry curvature \cite{NiuBerry,Matsumoto,Matsumoto2,RShindou,RShindou2,RShindou3,Mook,Mook2,Mook3,Lifa,KevinHallEffect}
is a local quantity that reflects the geometric properties of the Bloch wavevector-space,
it can be expected that quantum Hall effects emerge also in magnonic systems.
A Landau level is \cite{QHEmagnon} formed also in Dirac magnon systems.
Taking that into account to exploit quantum Hall effects of Dirac magnons and to reveal their relativistic effects ({\it{e.g.}}, quantum anomaly) via magnon transport is certainly an interesting future direction \cite{SPTnote}.
%%%%%%%%%%%%%%%%%%%%%%%%%%%%%
A similar approach \cite{QHEweyl} will be useful to investigate the transport properties of {\it{Weyl magnons}} \cite{WeylMagnon}.
Ref. [\onlinecite{MookWeylMagnon}] demonstrated that magnonic Weyl points can be controlled by external magnetic fields.

%%%%%%%%%%%%%%%%%%
\subsection{Optomagnonics}
\label{subsec:opmag}
%%%%%%%%%%%%%%%%%%

With the recent demonstration of the  coherent coupling to a superconducting qubit in Ref. [\onlinecite{tabuchiScience}],
hybrid structures involving the coherent coupling between insulating magnets with photonic \cite{Optomagnonics,Optomagnonics2,Optomagnonics3,Optomechanics} and electronic degrees of freedom has certainly a bright future.

\begin{acknowledgments}
We (KN and DL) acknowledge support by the Swiss National Science Foundation and the NCCR QSIT and by the JSPS (KN: Fellow No. 26-143).
Many results reviewed in this article have been obtained in collaboration with a number of colleagues over the past years. 
In particular, it is a special pleasure to express our thanks to Kevin van Hoogdalem, Jelena Klinovaja, and Yaroslav Tserkovnyak for their valuable contributions.
\end{acknowledgments}

%%%%%%%%%%%%%%%
%%%%%%%%%%%%%%%
%%%%%%%%%%%%%%%

%\appendix

\bibliography{PumpingRef}

\begin{thebibliography}{240}
\bibitem[{BEC({\natexlab{a}})}]{BECcurrentNOTE}
\bibinfo{note}{An experimental realization of `a magnon-supercurrent'
  \cite{katsura2,sonin,BrataasSuperfluid} was reported in Ref.
  [\onlinecite{MagnonSupercurrent}]. See Ref. [\onlinecite{LeggettRevHe}] for
  superfluid in liquid $^{3}\rm{He}$.}
\bibitem[{NAN()}]{NANDnote}
\bibinfo{note}{NAND gate, one of the universal gates for classical computation,
  is a two-bit gate that provides a logical $0$ as outcome only if both the
  input bits are $1$, while it yields a logical $1$ otherwise.}
\bibitem[{Mag()}]{MagnonicQHEnote}
\bibinfo{note}{In Ref. [\onlinecite{QHEmagnon}], providing a topological
  description \cite{NiuBerry,Kohmoto,TKNN} of the classical magnon Hall effect
  \cite{magnon2} induced by the A-C phase, we discuss the condition for
  magnonic `quantum' Hall conductivities characterized by Chern number
  associated with Berry curvature of Bloch wavevector and provide the condition
  for the magnonic WF law \cite{magnonWF} to demonstrate the universality. See
  Ref. [\onlinecite{Haldane2}] for the magnonic quantized Hall conductivity by
  phase twist\cite{NiuPhaseTwist}.}
\bibitem[{BEC({\natexlab{b}})}]{BECnoteLeggett}
\bibinfo{note}{A macroscopic number of magnons occupies a single or several
  states, which is called single or fragmented condensation, respectively. See
  Refs. [\onlinecite{leggett,LeggettReview}] for details.}
\bibitem[{Bas()}]{BassoNOTE}
\bibinfo{note}{Using Johnson-Silsbee approach \cite{SilsbeeMagnetization} and
  the thermodynamics out-of-equilibrium, Refs. [\onlinecite{Basso,Basso2}] have
  microscopically introduced a magnon chemical potential that takes a nonzero
  value only in nonequilibrium.}
\bibitem[{not({\natexlab{a}})}]{noteJosephson}
\bibinfo{note}{We note this is a rough estimation. More microscopic evaluation
  will be desired.}
\bibitem[{SPT()}]{SPTnote}
\bibinfo{note}{As in Ref. [\onlinecite{SenthilLevin}], such approach from
  symmetry-protected topological phase \cite{SPTtextWen} might be useful.}
\end{thebibliography}

\end{document}